# pH-dependent water permeability switching and its memory in 1T' MoS$_2$ membranes

C. Hu[1,2,3,4], A. Achari[1,2], P. Rowe[5], H. Xiao[1,2], S. Suran[1,2], Z. Li[6], K. Huang[1,2], C. Chi[1,2], C. T. Cherian[1,2,7], V. Sreepal[1,2], P. D. Bentley[8], A. Pratt[8], N. Zhang[1,2,6], K. S. Novoselov[3,9], A. Michaelides[5], R. R. Nair[1,2]

[1]National Graphene Institute, University of Manchester, Manchester, M13 9PL, UK.
[2]Department of Chemical Engineering, University of Manchester, Manchester, M13 9PL, UK.
[3]Department of Physics and Astronomy, University of Manchester, Manchester M13 9PL, UK.
[4]College of Chemistry and Chemical Engineering, iChEM, Innovation Laboratory for Sciences and Technologies of Energy Materials of Fujian Province (IKKEM), Xiamen University, Xiamen, China.
[5]Yusuf Hamied Department of Chemistry, University of Cambridge, Cambridge CB2 1EW, UK.
[6]School of Chemical Engineering, Dalian University of Technology, Panjin, Liaoning, 124221, China.
[7]Department of Physics and Electronics, CHRIST (Deemed to be University), Bangalore, 560029, India.
[8]Department of Physics, University of York, York, YO10 5DD, UK.
[9]Centre for Advanced 2D Materials and Graphene Research Centre, National University of Singapore, 117546 Singapore.
amritroop.achari@manchester.ac.uk, rahul@manchester.ac.uk

**Intelligent transport of molecular species across different barriers is critical for various biological functions and is achieved through the unique properties of biological membranes[1-4]. An essential feature of intelligent transport is the ability to adapt to different external and internal conditions and also the ability to memorise the previous state[5]. In biological systems, the most common form of such intelligence is expressed as hysteresis[6]. Despite numerous advances made over previous decades on smart membranes, it is still a challenge for a synthetic membrane to display stable hysteretic behaviour for molecular transport[7-11]. Here we show the memory effects and stimuli regulated transport of molecules through an intelligent phase changing MoS$_2$ membrane in response to external pH. We show that water and ion permeation through 1T' MoS$_2$ membranes follows a pH dependent hysteresis with a permeation rate that switches by a few orders of magnitude. We demonstrate that this phenomenon is unique to the 1T' phase of MoS$_2$ due to the presence of surface charge and exchangeable ions on the surface. We further demonstrate the potential application of this phenomenon in autonomous wound infection monitoring**

and pH-dependent nanofiltration. Our work significantly deepens understanding of the mechanism of water transport at the nanoscale and opens an avenue for developing neuromorphic applications, smart drug delivery systems, point-of-care diagnostics, smart sensors, and intelligent filtration devices.

One possible direction to achieve an intelligent membrane is the use of phase-changing materials, which can modify their structure depending on external conditions. Two-dimensional transition metal dichalcogenides (TMDCs) such as $MoS_2$ is a strong candidate for this since it can exist in (and can be controllably switched between) several structural phases (Fig. 1a-c)[12-15]. The most stable polytype of bulk $MoS_2$ is the hexagonal 2H phase. $MoS_2$ also exists as many metastable trigonal forms, such as 1T, 1T', and 1T'', distinguished by a series of Peierl's distortion, resulting in the formation of different superstructures[14]. Among these polytypes, 1T' is the energetically preferred metastable phase with an excess negative charge in which Mo atoms are clustered into zigzag chains[16]. Recently, the potential of TMDC membranes for water filtration applications has been investigated, and fast water transport through pristine and functionalised $MoS_2$ membranes has been observed[17-25].

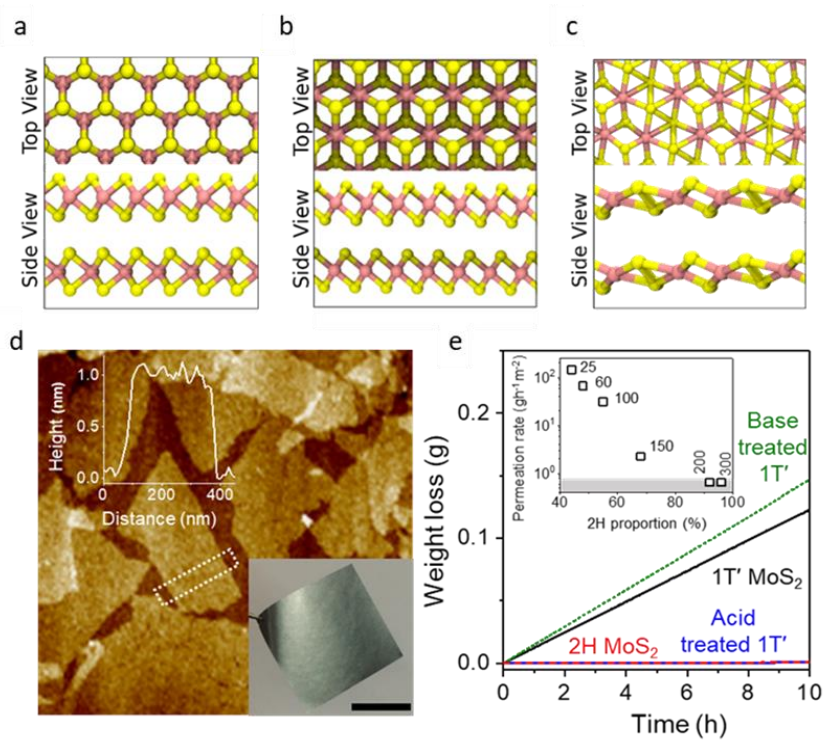

*Fig. 1| Water permeation through $MoS_2$ membranes. (a-c)* Schematic models of the 2H, 1T and 1T' phase of $MoS_2$ in top and side view. *(d)* AFM image of $MoS_2$ flakes with height profile (top inset) along the dotted rectangle. Bottom inset: Photograph of a 2 µm thick freestanding $MoS_2$ membrane. Scale bar, 1 cm. *(e)* Weight loss for a water filled container sealed with as-prepared 1T', 2H, acid (0.1 M HCl) and base (0.1 M LiOH) treated (soaked) $MoS_2$ membranes with a thickness of 2 µm (aperture ≈ 1 $cm^2$) (colour coded labels). Inset: variation of water vapour permeation rate as a function of proportion of 2H phase in the $MoS_2$ membrane. The proportion of 2H phase (as determined from the XPS analysis) in the $MoS_2$ membrane is tuned by annealing the membrane at different temperatures in an inert atmosphere. Temperatures (in degree Celsius) at which the membrane was annealed are indicated on the graph. Grey area shows the below-detection limit for our measurements.

Our MoS$_2$ membranes (Fig. 1d inset) were prepared by lithiation of bulk MoS$_2$ powder (2H phase) followed by exfoliation and vacuum filtration of the dispersion as reported previously (Materials and method)[18]. Atomic force microscopy (AFM) studies reveal that the dispersion primarily consists of single-layered MoS$_2$ sheets with a thickness of ≈1 nm (Fig. 1d). It is known that exfoliation of 2H MoS$_2$ by lithiation leads to massive electron doping into MoS$_2$, which induces a phase transition from the 2H phase to the 1T' phase[14]. Accordingly, X-ray photoelectron spectra (XPS) of our as-prepared MoS$_2$ membranes (Supplementary Fig.1) reveal that the primary phase is 1T' (56%) (Supplementary section 1), in agreement with previous reports[18]. Since 1T' is the dominant phase in the as-prepared membranes, hereafter we refer to them as 1T' MoS$_2$ membranes. These membranes can be converted back to the 2H phase by annealing them at 300 °C in inert atmosphere (Supplementary Fig. 1)[13].

Water vapour permeation properties of 1T' and 2H MoS$_2$ membranes were studied by weight loss measurements from a water-filled metal container sealed with 2 μm thick MoS$_2$ membranes as reported previously (Materials and method)[26]. We observed a significant weight loss for the 1T' MoS$_2$ membrane, but no weight loss for the 2H MoS$_2$ membrane (Fig. 1e). In addition, increasing the proportion of 2H phase in the 1T' MoS$_2$ membranes by controlled annealing resulted in a monotonic decrease in water permeation; when the 2H proportion reached above 90%, the water permeation dropped below the detection limit of our experiment (inset of Fig. 1e). To rule out the effect of membrane drying during phase conversion on the permeation properties, we performed additional experiments to check the water permeation of vacuum-dried MoS$_2$ membranes (Supplementary section 2). We found that although vacuum drying removed intercalated water, the membrane soon recovered to a hydrated state after exposure to ambient conditions and allowed the permeation of water (Supplementary Fig. 2 and 3). To further explore water permeation through MoS$_2$ laminate membranes, we performed water vapour permeation experiments after treating the MoS$_2$ membranes in acidic and basic aqueous solutions with different pH. For this, the membranes were first soaked in acidic (HCl) or basic (LiOH) solutions of the desired pH (ranging from pH 1 to pH 12.1) for 10 h, followed by washing with deionised water and drying. To our surprise, we found that 1T' MoS$_2$ membranes treated at pH 1 (0.1 M HCl) block water vapour permeation (Fig. 1e and Fig. 2a) whereas after exposing the same membrane to pH 12.1 (0.1 M LiOH), it recovered its water vapour permeation to the same level of the as-prepared 1T' MoS$_2$ membranes.

Fig. 2a shows the variation in water vapour permeation rate of 1T' and 2H MoS$_2$ membranes with pH during a full cycle of pH change (acidic to basic and back to acidic). Interestingly, while increasing the pH of the soaking solution from 1 to 12.1, the membrane remains impermeable until it is treated with a solution of pH ~11, after which it rapidly switches from a non-permeating state to a permeating state. Similarly, while decreasing the pH of the soaking solution from 12.1 to 1, the membrane remains water permeable until it is treated with a solution of pH ~4, at which point it rapidly switches from a permeable to an impermeable state (Fig. 2a). This hysteresis behavior of water vapour permeation switching is unexpected. In comparison, the 2H MoS$_2$ membranes remained impermeable throughout this pH range. This pH-controlled water permeation of 1T' MoS$_2$ membranes is reversible and robust. No degradation in membrane performance (inset of Fig. 2a) was observed even after ten pH cycles between 1 and 12.1. Additionally, to confirm the stability of the membrane performance we have conducted water vapour permeation after soaking acid and base treated membranes in deionised water for 48 hours and observed no change in their performance (Supplementary section 3, Supplementary Fig. 4). The observed pH-responsive behaviour of 1T' $MoS_2$ membranes was also confirmed while the membrane is in contact with different pH solutions (Supplementary section 4), hence the change in pH of the feed solution spontaneously alters the permeation rate (Supplementary Fig. 5). To understand the kinetics of the pH responsiveness, we studied the permeation as a function of pH exposure time (Supplementary section 5) and found that the response time of the membrane can be tuned by changing the thickness of the membrane. For example, for a membrane with a thickness of 500 nm, the response time was 2 min, while it needed 30-60 minutes for a membrane with 2 µm thickness (Supplementary Fig. 6).

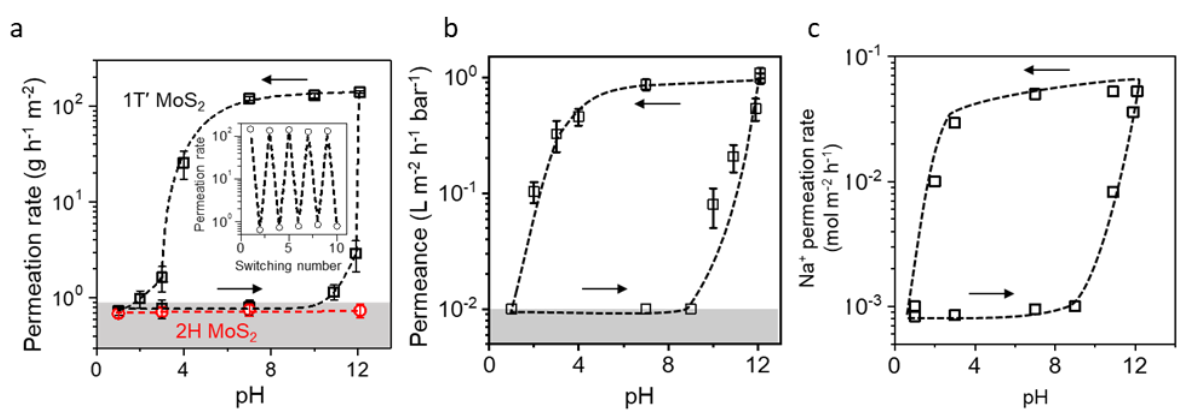

*Fig. 2| pH controlled water and ion transport through a 1T' $MoS_2$ membranes. (a) Water vapour permeation rate of 2H and 1T' $MoS_2$ membranes as a function of pH treatment. Dashed line is a guide to the eye and arrows indicate the direction of pH change (colour coded labels). Grey area shows the below-detection limit for our measurements. Inset: Reversible change in water permeation by alternating acid (0.1 M HCl) and base (0.1 M LiOH) treatment. (b) Liquid water permeance through a 1T' $MoS_2$ membrane (area of 1 $cm^2$ and thickness of 0.5 µm) as a function of pH treatment. The filtration experiments were performed at 5 bar pressure. Grey area shows the below-detection limit for our measurements. Dashed line is a guide to the eye and arrows indicate the direction of pH change. Both water vapour and liquid permeation through as-prepared fresh $MoS_2$ membranes were equal to that of pH 7 treated membranes. Error bars in **a** and **b** denote standard deviations using three different samples. (c) $Na^+$ ion permeation rate though a 2 µm thick pristine 1T' $MoS_2$ membrane as a function of pH of the 1 M NaCl feed solution. The permeate compartment was pure water. Dashed line is a guide to the eye and arrows indicate the direction of pH change.*

In accordance with the water vapor permeation, a pH-responsive hysteresis in liquid water permeation was also observed in 1T' $MoS_2$ membranes (Fig. 2b) in a pressure filtration system. Within our sensitivity of 0.01 L $m^{-2}$ $h^{-1}$ $bar^{-1}$, no water permeation was obtained for acid treated samples, whereas the permeation increased to 1.1 L $m^{-2}$ $h^{-1}$ $bar^{-1}$ after treating the same membrane with an alkali solution of pH 12.1. The pure water flux was also tested continuously for up to 40 hours and no significant changes in water flux were noticed, confirming the stability of the membranes (Supplementary section 6, Supplementary Fig. 7). Furthermore, we also tested the permeation of 0.1 M LiOH (pH 12.1) and 0.1 M HCl (pH 1) solution directly through $MoS_2$ membranes and found that the permeance can be repeatedly switched on or off depending on the pH of the feed solution (Supplementary Fig. 7).

To further demonstrate the potential of the unique pH-sensitive 1T' MoS$_2$ membranes, we performed two additional experiments (Supplementary section 7, 8). First, we tested ion permeation properties as a function of pH of the feed and pH of the MoS$_2$ membrane treatment (Supplementary section 7). Our results show that, similar to water permeation switching, ion permeation through 1T' MoS$_2$ membranes can also be switched by the pH of the feed solution or pH of the MoS$_2$ membrane treatment (Fig. 2c, Supplementary Fig. 8). Fig. 2c further shows that ion permeation also demonstrates hysteretic behaviour depending on the direction of change in the pH of the feed solution. Additionally, we studied the pH-dependent nanofiltration properties of MoS$_2$ membranes and demonstrated that the salt rejection of the membranes could be controlled by the pH of the membrane treatment (Supplementary Fig. 9). Second, we show that the pH responsiveness of MoS$_2$ membranes can be potentially used as a pH sensor for detecting infection in wounds (Supplementary section 8). Wound healing is a complex physiological process that requires continuous monitoring and attention. When a wound is infected, the pH of the wound increases from 4-5 (normal skin pH) to approximately 7-8 (chronic wound pH). Using simulated body fluid, we show that the MoS$_2$ membranes can be used to monitor the wound infection by autonomously detecting the pH (Supplementary Fig. 10) i.e. only when there are natural signs of infection.

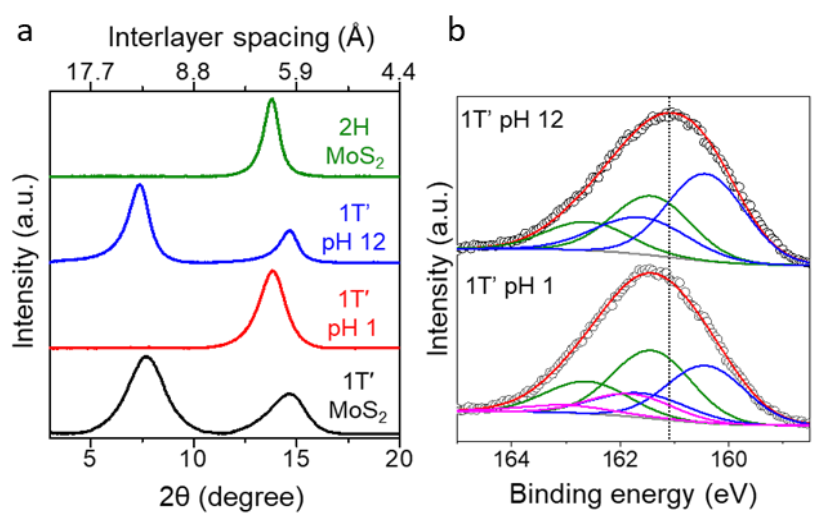

*Fig. 3 | Characterisation of MoS$_2$ membranes. (a) X-ray diffraction pattern of 1T', 2H, acid and base treated 1T' MoS$_2$ membranes (colour coded labels). (b) S 2p XPS spectra of acid and base treated 1T' MoS$_2$ membranes. Black circles, raw data; red line, the fitting envelope; blue lines and green lines, deconvolved peaks attributed to the 1T' and 2H phases, respectively. Magenta line, peak fitting attributed to S-H bonds. The dotted line is a guide to the eye for an overall shift in the envelope of the spectra by 0.5 eV.*

It has been generally accepted that permeation through 2D laminar membranes occurs through interplanar spaces and the interlayer distance plays an important role in governing the molecular permeation. As expected, the X-ray diffraction (XRD) pattern of 1T' MoS$_2$ shows two peaks at 2θ ≈ 7.7° (001 reflection) and 14.6° (002 reflection), corresponding to an interlayer spacing of 11.4 Å, whereas in the case of 2H MoS$_2$ the peak is located at 2θ ≈ 13.8° (002 reflection) corresponding to the interlayer spacing of 6.4 Å (Fig. 3a). The larger interlayer spacing in the 1T' phase is attributed to the presence of two layers of intercalated water (due to the humid ambient air)[27], which is further confirmed by X-ray analysis under vacuum where the interlayer distance decreases (Supplementary Fig. 2) and AIMD simulations (Fig. 4). Interestingly, the XRD pattern of the acid-treated 1T' $MoS_2$ membrane resembles that of the 2H $MoS_2$ membrane, however, the original spectrum is recovered after treating the membrane again with an alkaline solution. When the pH of the soaking solution was gradually decreased from 12 to 1, the (001) reflection of the 1T' $MoS_2$ membrane at 7.7° gradually decreases and finally disappears at pH 2 (Supplementary section 9, Supplementary Fig. 11). From these experiments, it is apparent that the water permeation follows the changes observed in the interlayer spacing; 2H and acid-treated 1T' $MoS_2$ membranes with a smaller interlayer spacing (6.4 Å) block water permeation, while pristine 1T' $MoS_2$ and base treated 1T' $MoS_2$ membranes with a larger interlayer spacing (11.4 Å) allow water to permeate.

Based on the smaller interlayer spacing, the absence of water permeation through 2H and acid-treated samples is not surprising. However, the reversibility of the interlayer spacing and pH-dependent hysteresis of water permeation through the 1T' $MoS_2$ membranes is intriguing. One possible explanation for the observed reversible change in the interlayer spacing and the associated water permeation is the reversible phase change of the $MoS_2$ membrane from the 1T' to the 2H phase. However, our XPS analysis on the acid and base treated samples rules out this possibility (Fig. 3b, Supplementary section 10). The only minor difference in the Mo 3d XPS spectra of the acid-treated samples is the slight decrease in the 1T' proportion from 56% to 46% (Supplementary Fig. 12). We do find, however, that after the acid treatment the S 2p peak shows a significant shift (to 0.5 eV higher binding energy) compared to the base treated 1T' $MoS_2$ membranes indicating electron transfer from the electron-rich 1T' phase. To obtain self consistently the same proportion of 1T' phase from both the Mo 3d and S 2p peak fitting, we had to include another peak at 161.8 eV (Fig. 3b), which corresponds to an S-H peak (16%), suggesting the binding of a proton to the sulfur atoms of $MoS_2$ in the acidic conditions, as predicted previously[28,29]. To further probe the influence of pH on the atomic structure of 1T' $MoS_2$, we have conducted X-ray absorption spectroscopy (XAS) measurements (Supplementary section 11) and found significant changes in bond coordination numbers after acid treatment (Supplementary Figs. 13, 14), suggesting partial and reversible changes in the atomic structure after acid treatment.

It is known that the 1T' phase of $MoS_2$ is relatively unstable and can be stabilised by the intercalation of alkali metal ions[15,30]. Since the studied pristine 1T' membranes were prepared by lithiation and the base treatment involves LiOH, we have carefully analysed the Li content in the studied 1T' $MoS_2$ membranes by using inductively coupled plasma mass spectrometry (ICP-MS). As-prepared pristine 1T' membranes show Li content up to 0.3 mmol/g $MoS_2$, but after acid and base treatment, the concentration of $Li^+$ ions changed significantly. Fig. 4a shows the Li/Mo molar ratio of the $MoS_2$ membranes as a function of pH treatment. Similar to the water permeability, the Li/Mo molar ratio also changes with pH treatment and exhibits a hysteresis behavior. Above and below a critical pH, the Li/Mo ratio switches from low to high and high to low respectively. The similar trend in the changes of water permeation and the Li/Mo ratio as a function of pH treatment suggests a correlation between Li content in the membrane and the water permeation rate. Fig. 4b shows the water permeation rate as a function of Li/Mo ratio, demonstrating that for Li/Mo > 0.1, the water permeation rate is high and that it sharply decreases for lower Li/Mo ratios. In addition to $Li^+$, we have looked into the influence of different intercalating cations on the water permeation properties of $MoS_2$ membranes (Supplementary section 12) and observed similar behaviour in almost every case (Supplementary Figs. 15, 16).

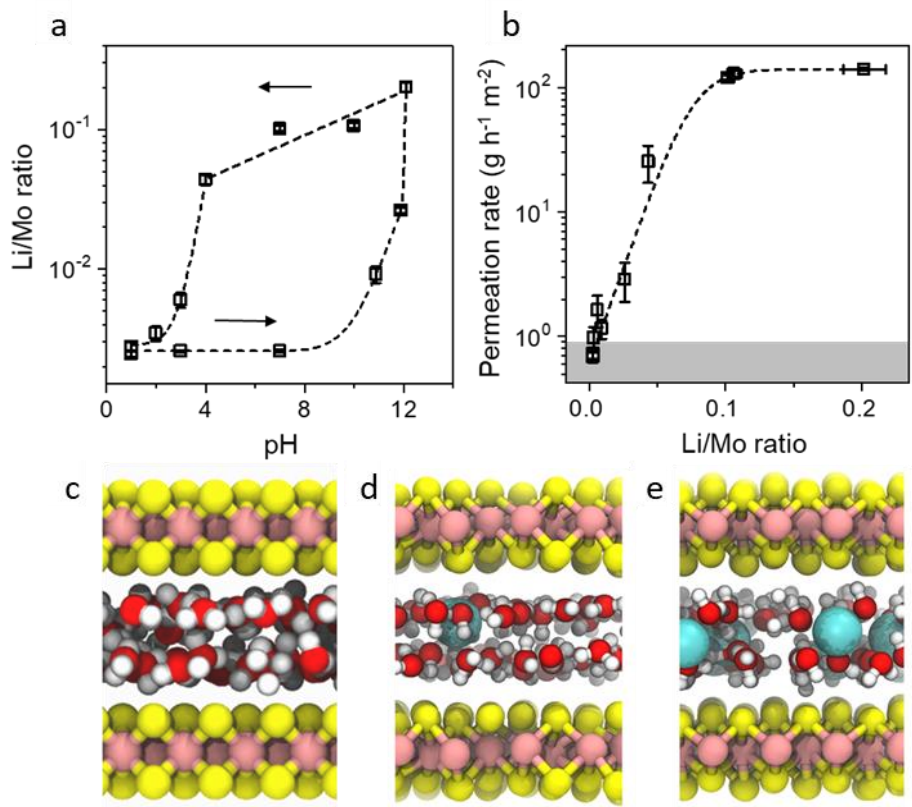

*Fig. 4| Influence of Li⁺ ions and MoS₂ phase on water permeation. (a)* Li/Mo ratio (as determined by ICP-MS) of MoS₂ membranes as a function of pH treatment. Dashed line is a guide to the eye and arrows indicate the direction of pH change. Since Li⁺ ions only interact with the negatively charged 1T' region, the Li/Mo ratio is estimated based on the Mo in the 1T' fraction. *(b)* Water permeation rate as a function of Li/Mo ratio. Dashed line is a guide to the eye. Grey area shows the below-detection limit for our measurements. Error bars in *a* and *b* denote standard deviations using three different samples. Snapshot from simulation of water bilayer structure confined between 2H *(c)*, 1T' MoS₂ with low (0.02)*(d)* and high (0.3)*(e)* Li/Mo ratio. Yellow, sulfur; pink, molybdenum; teal, lithium; red, oxygen; white, hydrogen.

To elucidate the mechanism of water permeation through MoS₂ membranes in a greater extent, we have performed water vapour adsorption isotherms on different MoS₂ samples (Supplementary section 13). Supplementary Fig. 17 shows the water uptake of an MoS₂ membrane as a function of the pH treatment. The water adsorption follows a typical type 2 adsorption isotherm and suggests monolayer and multilayer water adsorption between layers of MoS₂ followed by capillary condensation. The water uptake at 99% relative humidity for different pH treated samples exhibits a pH-dependent hysteresis (Supplementary Fig. 17), similar to that of water permeation and Li/Mo ratio. Based on the correlation between the Li⁺ ion adsorption and water permeation and water adsorption, we propose that water permeation through MoS₂ laminates is governed by an adsorption-diffusion mechanism[31], where water molecules are adsorbed on the cations by hydration, and then diffuse between the layers of MoS₂. During the acid treatment of the MoS₂ membranes, the adsorbed Li⁺ ions were exchanged with protons which interact strongly with the negatively charged 1 T' MoS₂ surface to form S-H bonds, as evidenced by our XPS data in Fig. 3b[28,29]. Due to this, the interlayer charges disappear and hence prevent water adsorption resulting in a smaller interlayer spacing and no permeation between the layers of MoS$_2$. Owing to the weak acidic character of S-H bonds, exposing the acid-treated samples to higher pH leads to bond cleavage and re-adsorption of cations into the MoS$_2$ layers. Since the pKa of S-H bond is high, deprotonation occurs only at high pH. This is further experimentally verified by treating acid-treated MoS$_2$ membranes with LiCl solution where no change in the permeation properties was noticed, suggesting proton to Li exchange only occurs at high pH (Supplementary section 14, Supplementary Fig. 18). In the reverse process, sufficient exchange of interlayer Li$^+$ with protons can only happen at high external H$^+$ ion concentration, i.e. at pH 4 or below. Hence switching of permeability occurs at high/low pH values depending on the direction of pH change. This proposed model explains the observed hysteresis in the permeation data and also supports the observed hysteresis in the Li$^+$ ion adsorption as a function of pH. We have also found that protonation/delithiation or deprotonation/lithiation follows an activated behavior (Supplementary section 15) with an activation barrier of 30 kJ/mol (Supplementary Fig. 19). To further validate this hypothesis, we exposed as-prepared 1T' MoS$_2$ membranes to hydrogen plasma (Supplementary section 16). The hydrogen plasma can hydrogenate the S atoms on the membrane surface. As expected, the hydrogenated membrane showed blockage of water vapour. The permeation was returned to its original value after re-exposing the membranes to LiOH solution. (Supplementary Fig. 20).

To further demonstrate the unique water permeation properties of MoS$_2$ membranes, we performed a control experiment (Supplementary section 17) with a polar solvent ethanol, which intercalates between MoS$_2$ layers (Supplementary Fig. 21) and has comparable adsorption to that of water (Supplementary Fig. 21). Surprisingly, we found that even though ethanol can be adsorbed into the membrane, the permeation rate of ethanol was two orders of magnitude lower than that of water (Supplementary Fig. 22). This suggests that the observed permeation of water through MoS$_2$ is due to its unique faster diffusion between the MoS$_2$ layers.

*Ab initio* molecular dynamics (AIMD) simulations provide more insights into the mechanism of water transport between the layers of MoS$_2$ (Supplementary section 18). In these simulations, we examined the structure of water confined between the 2H, 1T', and 1T' phase with low (0.02) and high (0.3) Li/Mo ratio, all at a fixed interlayer separation of ~1 nm (Figs. 4c-e and Supplementary Figs. 23-26). Our simulations show that irrespective of the polymorph or Li concentration, the confined water forms well-defined bilayer structures (Fig. 4c-e, Supplementary Fig. 23, 25). The main difference between 1T' MoS$_2$ with high and low Li/Mo ratio (base and acid treated membranes) lie in the orientation of the water molecules and position of the Li ions in the MoS$_2$ interlayer space. In the case of 1T' MoS$_2$ with a high concentration of Li-ions, we observed a reorientation of the water molecules to point H atoms towards the surfaces of the confining MoS$_2$ (Fig. 4e, Supplementary Fig. 25). Furthermore, the Li ions were found to localise between the two water monolayers, as compared to 1T' MoS$_2$ with a low Li-ion concentration, when they were found to localise close to the MoS$_2$ surface (Supplementary Fig. 24). However, for both cases, water structures are commensurate with the underlying structure of the MoS$_2$ (Supplementary Fig. 26) and have a hydrogen bond network in which water molecules bond primarily to one another. The observed high probability of a water molecule being located directly above a S atom of the MoS$_2$ medium, compared with the low probability of a water molecule being located in the space between atoms is indicative of a hopping mechanism of water diffusion. To understand whether these tiny changes in water structure cause any difference in the dynamic properties of water molecules, we monitored the mean squared displacement (MSD) for water oxygen atoms and found that there is very little/no effect on the dynamics of water between low (Li/Mo= 0.02) and high (Li/Mo= 0.3) Li concentration (Supplementary Fig. 27). This suggests that both acid and base treated 1T′ $MoS_2$ channels allow fast water permeation due to the hydrogen-bonded water network and a relatively weak $MoS_2$-water interaction.

To further understand the difference in water permeation between 2H, acid and base treated 1T′ $MoS_2$ laminates, we have performed additional classical MD simulations to obtain insight into water entry effects (Supplementary section 19). Our simulation found that $Li^+$ ions adsorbed in the interlayers of 1T′ $MoS_2$ allow water to enter inside the channel spontaneously (Supplementary Fig. 28) (swelling). We propose that this spontaneous swelling originates from the osmotic effects of ions between the layers of $MoS_2$ (osmotic pressure driven water intercalation). In the absence of ions (2H and acid-treated $MoS_2$), the water molecules did not enter the channels (Supplementary Fig. 28-30). This explains why 2H membranes and acid-treated $MoS_2$ membranes are impermeable to water while 1T′ is permeable. The data in the Fig. 1e inset also supports this model, where the water permeation rate exponentially decreases with increasing 2H fraction. Due to the absence of spontaneous water intercalation in 2H regions, they act as random resistance/blockades for water transport and hence with growing their numbers, water permeation decreases exponentially.

In conclusion, we demonstrate that 1T′ $MoS_2$ laminate membranes are permeable to water while 2H $MoS_2$ laminate membranes are impermeable to water. The presence of charge/ions between the interlayers of 1T′ $MoS_2$ membranes facilitates water adsorption into the membrane by cation hydration. The absence of any charge between the layers of 2H $MoS_2$ explains its water impermeability. The charge in the 1T′ $MoS_2$ layers is controlled by pH, resulting in the switching of water permeation; at acidic pH, the S atoms go through reversible hydrogenation, removing the charges from the layers and cations from the interlayer spaces, and hence the membranes become impermeable. While at basic pH, both the adsorbed cation concentration and water permeation recover. The observed hysteresis in water and ion permeation with pH is explained by the extreme pH required for protonation and deprotonation of $MoS_2$. The fast water transport through $MoS_2$ membranes is found to be independent of the nature of the $MoS_2$ layers, rather, it is due to the hydrogen-bonded network of the confined water and the relatively weak water-$MoS_2$ interactions. Our work demonstrates a pH-responsive intelligent membrane based on $MoS_2$ laminates, and also deepens understanding of the mechanism of water permeation through 2D-material-based-membranes.

**Materials and Methods:**
Exfoliation of $MoS_2$.
$MoS_2$ was exfoliated by the lithium intercalation method as reported previously[19]. Briefly, 0.5 g $MoS_2$ powder (purchased from Sigma-Aldrich) was mixed with n-butyllithium (5 mL 1.6 M in hexane purchased from Sigma-Aldrich) and stirred inside a glovebox for 48 h. The mixture was then filtered and washed with hexane three times. The intercalated powder was washed twice with water and then sonicated for 1 h in water and centrifuged at 12000 rpm to remove non-exfoliated materials.

Preparation of MoS$_2$ membranes.

MoS$_2$ membranes were fabricated by vacuum filtration of the MoS$_2$ suspension in water through a polyethersulfone (PES) membrane. We used PES membranes with 0.22 μm pore size (purchased from Millipore) to make MoS$_2$ membranes with a thickness of 2 μm for pervaporation measurements and PES with 0.03 μm pore size (purchased from Sterlitech) to make MoS$_2$ membranes with a thickness of 0.5 μm for pressure filtration measurements. Thermal annealing was carried out using MoS$_2$ membrane on an anodic alumina membrane with a pore size of 0.2 μm at 60 to 300 °C for 1 h under argon atmosphere. Acid and base treatments were conducted by soaking the as-prepared membranes in 0.1 M HCl (pH 1) or 0.1 M LiOH (pH 12.1) for 10 hours, then washed with deionized water and dried at room temperature. Aqueous solutions with different pH were prepared by diluting 0.1 M HCl and 0.1 M LiOH. The pH of diluted solutions was measured using a pH meter (Seven Excellence, Mettler Toledo).

Water vapour permeation measurements.

Permeation properties of the various MoS$_2$ films were measured using techniques that were described in detail previously. In brief, for vapour permeation measurements, dry as-prepared MoS$_2$ membranes with a thickness of 2 μm on PES substrates were glued to a Cu foil using Master Bond (EP41S-5) epoxy resin with an opening area of 1 cm$^2$. The foil was clamped between two nitrile rubber O-rings sealing a metal container. Permeability was measured by monitoring the weight loss of the container that was filled with water inside a glovebox.

Pressure filtration measurements.

Liquid water pressure filtration was carried out using a dead-end filtration set up (Sterlitech HP4750) under a pressure of 5 bar with a membrane thickness of 0.5 μm and an area of 1 cm$^2$. The water flux was acquired after the membrane reached a steady-state (~2 hours). It should be noted that the dry membrane was first soaked in 0.1M LiOH to wet the membrane, then tested in a wet state after soaking in different pH solutions since it has been reported that a completely dried MoS$_2$ membrane has no liquid water flux[19].

Characterization.

XRD measurements were performed using a Rigaku SmartLab XRD system with Cu Kα radiation (λ = 0.154 nm). AFM was measured using a Bruker Dimension FastScan in the tapping mode under ambient conditions. ICP-MS was measured by fully decomposing 10 mg MoS$_2$ membrane in aqua regia at 60 °C, and then diluting with DI water.

**Acknowledgements:**

This work was supported by the Royal Society, the Leverhulme Trust (PLP-2018-220), the Engineering and Physical Sciences Research Council (EP/P00119X/1), Graphene Flagship, Carlsberg Research Laboratory, and European Research Council (contract 679689). C. H. acknowledges support from the China Scholarship Council. We thank the XAFS station (BL14W1) of the Shanghai Synchrotron Radiation Facility and R. Qin, W. Zhang for the XAFS measurement. We thank V. G. Kravets, Y. Su and K. -G. Zhou for discussions. N. Z. acknowledges support from the National Key Research and Development Program (2021YFC2901300). A.M. acknowledges support from the European Research Council under the European Union's Seventh Framework Programme (FP/2007-2013) / ERC Grant Agreement No. 616121 (HeteroIce project). We are grateful for computational support from the UK Materials and Molecular Modelling Hub, which is partially funded by EPSRC ((Grant Nos. EP/P020194/1 and EP/T022213/1), for which access was obtained via the UKCP consortium and funded by EPSRC grant ref EP/P022561/1. The authors acknowledge the use of the facilities at the Henry Royce Institute and UCL Grace High Performance Computing Facility (Grace@UCL), and associated support services.

# Supplementary information

1. X-ray photoelectron spectroscopy (XPS) of MoS$_2$ membranes

XPS spectra were acquired in an ultrahigh vacuum system with a base pressure of < 3 × 10$^{-10}$ mbar using a monochromated Al Kα source at 1486.6 eV (Omicron XM 1000) and a power of 220 W. An aperture diameter of 2 mm was used with the sample normal at 45° to both the X-ray source and the entrance optics of the hemispherical energy analyser (Omicron EA 125).

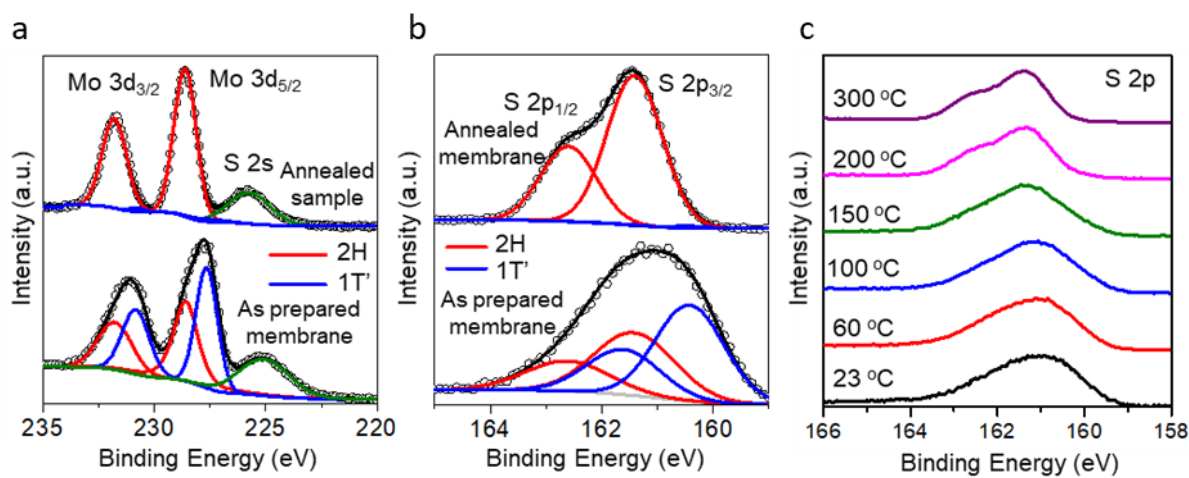

**Supplementary Fig. 1. X-ray photoelectron spectroscopy (XPS) of MoS$_2$ membranes.** XPS spectra of (**a**) Mo 3d, and S 2s states and (**b**) the S 2p region of as-prepared and 300 °C annealed MoS$_2$ membranes. Black circles, raw data; black line, the fitting envelope; blue lines and red lines, deconvolved peaks attributed to the 1T' and 2H phases, respectively. Green line, peak fitting for S 2s peak. (**c**) XPS spectra of S 2p region from a MoS$_2$ membrane annealed at different temperatures under argon atmosphere.

Supplementary Figs. 1a, b shows Mo 3d, S 2s, and S 2p regions of XPS spectra of an as-prepared MoS$_2$ membrane and the membrane after annealing at 300 °C in an inert atmosphere. To estimate the proportion of 1T' and 2H phases, we deconvoluted the Mo 3d and S 2p peaks since it is known that the peak corresponding to the 1T' phase is downshifted by ≈ 0.9 eV compared to the 2H phase[1]. The Mo 3d spin-orbit doublet with Mo 3d$_{3/2}$ and Mo 3d$_{5/2}$ components was fitted using a fixed intensity ratio of 2/3 and a doublet separation of 3.15 eV. The S 2p peak was fitted with S 2p$_{1/2}$ and S 2p$_{3/2}$ components using a fixed intensity ratio of 0.5 and a doublet separation of 1.2 eV. Deconvolution of the Mo 3d and S 2p regions shows that the concentration of the 1T' phase reaches ≈56% in the as-prepared samples, whereas the 300 °C annealed sample is completely converted into the 2H phase, in agreement with previous reports[1]. Supplementary Fig. 1c shows the evolution of the S 2p peak of an MoS$_2$ membrane from room temperature to 300 °C. It is apparent that the proportion of 2H phase increases with increasing temperature and 2H phase becomes predominant after annealing above 200 °C.

2. The effect of membrane drying on water permeation

To study the influence of membrane drying on the water permeation, we performed X-ray diffraction (XRD) and water permeation through MoS$_2$ membranes before and after exposing them to a vacuum. XRD at vacuum was performed using an Anton Paar DHS 1100 hot stage under 0.05 mbar pressure at room temperature after keeping the sample in a vacuum for 10 hours. The shift of the 001-diffraction peak under vacuum from 7.7° to 13° and its recovery upon exposing the membrane to air (Supplementary Fig. 2) clearly demonstrates that although vacuum drying removed intercalated water, the membrane soon recovered to a hydrated state after being exposed to the ambient conditions due to the strong hydration ability of Li$^+$. Thus, no significant changes in water vapour permeation rate were observed from the vacuum-dried membrane compared to an as-prepared membrane (Supplementary Fig. 3). It is to be noted that we do not observe any liquid water permeation for vacuum treated membranes in a pressure filtration setup, however, the membrane recovers its permeability after base treatment.

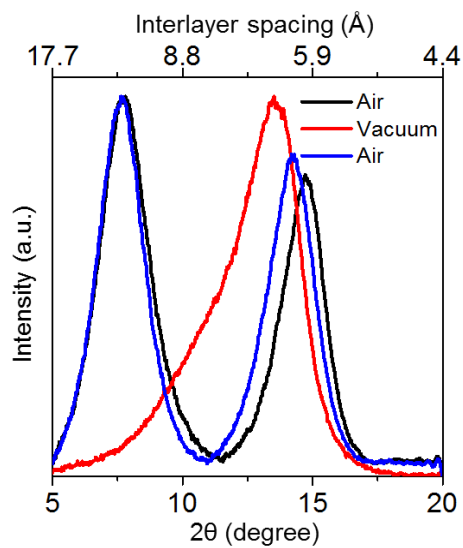

**Supplementary Fig. 2. X-ray diffraction of 1T' MoS$_2$ membrane under air and vacuum.** X-ray-diffraction of an as-prepared 1T' MoS$_2$ membrane under ambient conditions (black), vacuum (red) and after re-exposure to ambient condition (blue).

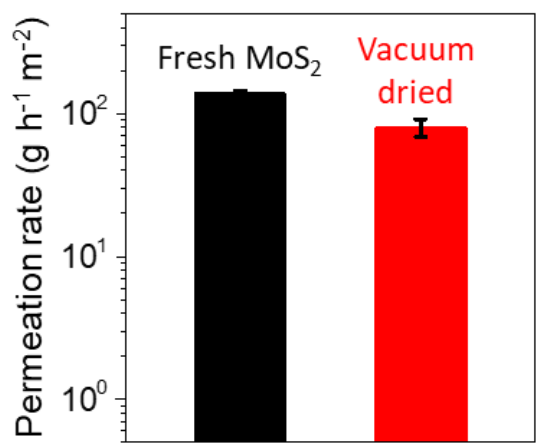

**Supplementary Fig. 3. The effect of membrane drying on water permeation.** Water vapour permeation rate of fresh and vacuum-dried MoS$_2$ membranes. Error bars denote standard deviations using three different samples.

3. Stability of acid and base treated 1T' MoS₂ membranes in deionized (DI) water

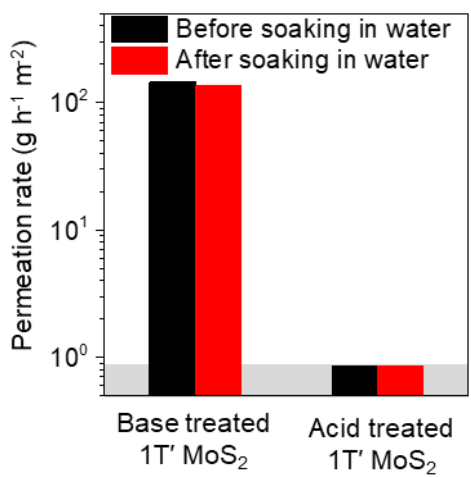

**Supplementary Fig. 4. Stability of acid and base treated 1T' MoS₂ membranes.** Water vapour permeation rate of acid and base treated 1T' MoS₂ membranes before and after soaking in DI water for 48 hours. Grey area shows the below-detection limit for our measurements.

4. pH-controlled pervaporation of water

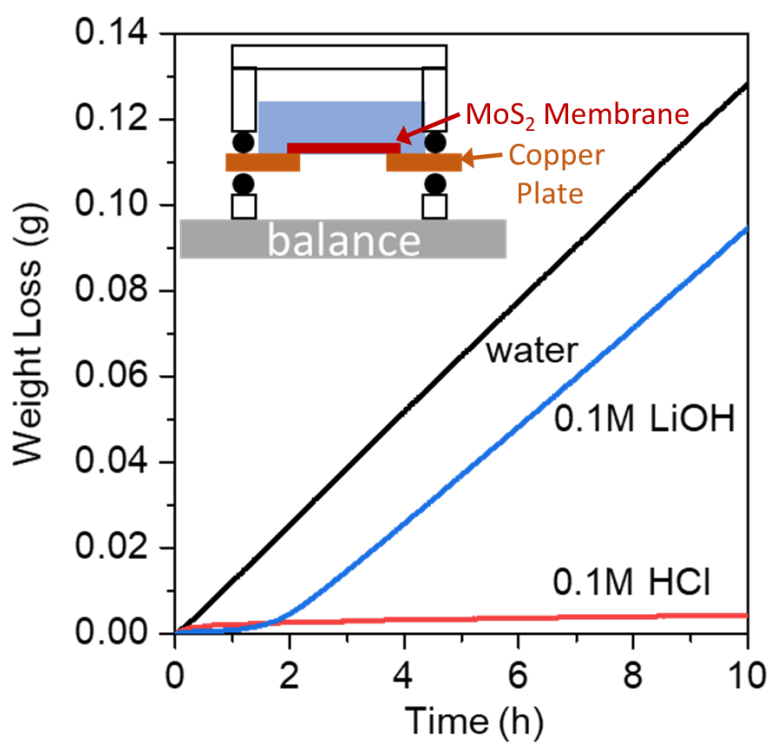

**Supplementary Fig. 5. pH-controlled water pervaporation through a 2 μm thick 1T' MoS₂ membrane.** Weight loss for a container filled with solutions with different pH in contact with 1T' MoS₂ membrane. When the container is filled with an acidic solution, no permeation was noted. However, after changing the solution to basic pH, the membrane switches back to a permeable state. Experiments were performed in the order of pH 1, 12.1 and 7. Inset: Schematic of our experimental setup.

To study the in-situ response of the 1T' MoS$_2$ membrane to acid and base solution, we performed pervaporation experiments where the membrane is in contact with the liquid. Supplementary Fig. 5 inset shows the schematic of the experimental setup. These experiments were performed by placing the liquid-filled containers upside down so that the membrane is in contact with the liquid. As shown in supplementary Fig. 5, when the container is filled with an acidic solution, no permeation was noted. However, after changing the solution to basic pH, the membrane switches back to a permeable state.

5. pH-response time of MoS$_2$ membranes

To study the pH response time of 1T' MoS$_2$ membranes, we performed water vapour permeation experiments as a function of membrane soaking time. Supplementary Fig. 6 shows that the pH-response time is highly dependent on the thickness of the membranes. For a membrane with 500 nm thickness, the water permeation rate decreased by 100 times after soaking in pH 1 solution for 2 min, while for a 2 μm thick membrane, it takes 30-60 minutes to achieve low permeability. Similarly, for a 2 μm acid-treated sample to recover its permeation, it takes ~30 minutes (Supplementary Fig. 6).

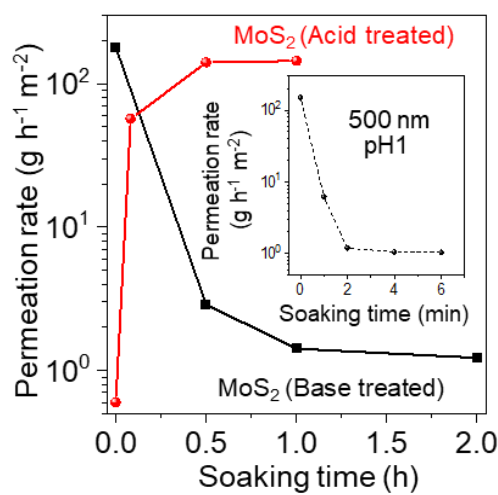

**Supplementary Fig. 6. pH-response time of 1T' MoS$_2$ membranes**. Water permeation rate as a function of soaking time in 0.1M HCl and 0.1 M LiOH solutions for both base treated and acid-treated 2 μm thick MoS$_2$ membranes, respectively. Inset: pH response time for a base treated 500 nm MoS$_2$ membrane as a function of soaking time in 0.1 M HCl.

6. Stability of 1T' MoS$_2$ membranes

To check the stability of the 1T' MoS$_2$ membrane and adsorbed Li$^+$ ions against liquid water permeation, we performed water filtration experiments for up to 40 hours. No significant reduction in water permeance suggests that the Li$^+$ ions and MoS$_2$ membranes are stable against water filtration. We also tested repeated acidic and basic solution filtration through 1T' MoS$_2$ membranes and found that permeance can be repeatedly switched on or off depending on the pH of the feed solution (Supplementary Fig. 7 inset).

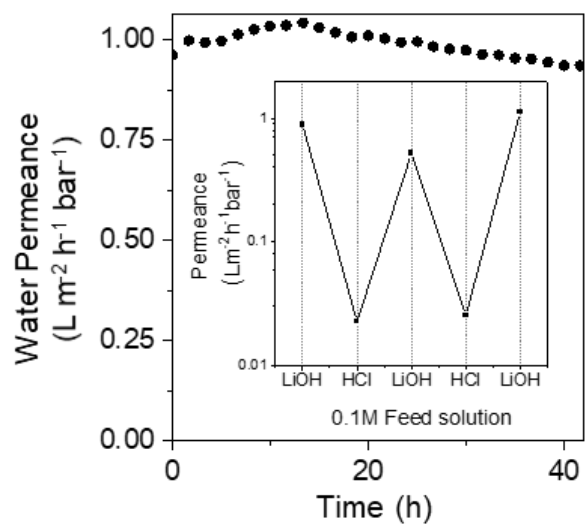

**Supplementary Fig. 7. Stability of 1T' MoS$_2$ membrane:** Liquid water permeance under 5 bar pressure through a 500 nm thick 1T' MoS$_2$ membrane treated with pH 12.1 solution (0.1 M LiOH) continuously for 40 hours. Inductively coupled plasma mass spectrometry (ICP-MS) analysis of the permeate does not detect the presence of Li ions. Inset: The permeance of a 500 nm thick 1T' MoS$_2$ membrane during pressure filtration of 0.1 M basic and acidic solutions showing membrane reversibility. Pressure filtration experiments were performed using a polycarbonate stirred cell from Sterlitech at 5 bar pressure.

7. pH-dependent ion diffusion and nanofiltration through 1T' MoS$_2$ membranes

Our measurement setup for studying ion diffusion through 1T' MoS$_2$ membranes was similar to the one previously reported[2] and consisted of two Teflon compartments (feed and permeate) separated by a 1T' MoS$_2$ membrane. The feed and permeate compartments were filled with 10 mL of 1M salt solution (NaCl) and deionized water, respectively. Ion permeation through MoS$_2$ membrane is studied by monitoring the concentration of ions in the permeate compartment by measuring the conductivity of the permeate compartment and also by measuring the amount of ions permeated by using ICP-MS. Supplementary Fig. 8a shows permeate compartment conductivity as a function of time for a 1T' MoS$_2$ membrane and the same membrane subsequently treated by acid and base. In addition to 1T' MoS$_2$ membranes, we also studied the ion permeation through a 2H MoS$_2$ membrane. As expected, the pristine 1T' MoS$_2$ and base treated MoS$_2$ membranes were permeable to NaCl; however, the absence of change in permeate conductivity for acid-treated and 2H MoS$_2$ membrane suggests its impermeability toward NaCl solution. Acid treated membranes became permeable to Na$^+$ ions after treating them with 0.1 M LiOH. The reversible recovery of Na$^+$ ion permeation through a 1T' membrane is shown in Supplementary Fig. 8b by consecutive acid and base treatment. In addition to reversible switching of Na$^+$ ion permeation, it also follows a hysteresis in ion permeation (similar to water permeation) depending on the direction of change in the pH of the feed solution (Fig. 2c).

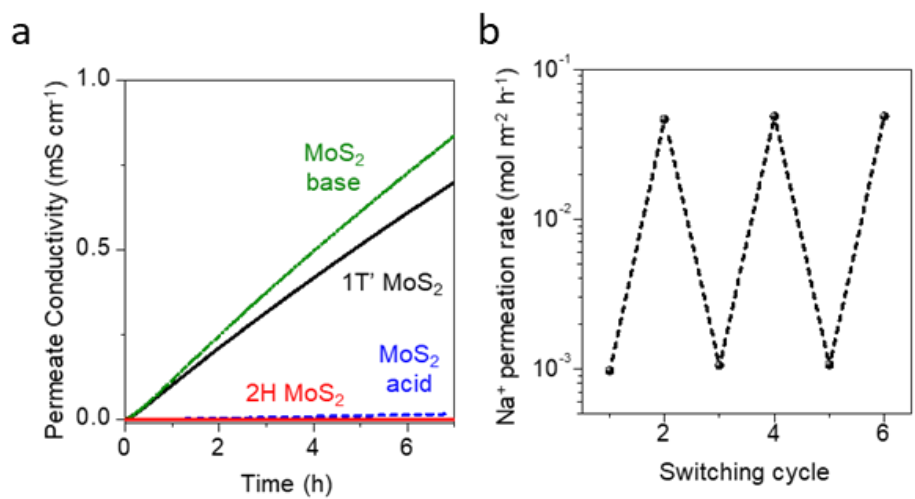

**Supplementary Fig. 8. pH-controlled ion diffusion through a 1T' MoS$_2$ membrane.** (a) Permeate conductivity as a function of time with 1 M NaCl and deionized water as feed and permeate side, respectively through 2 μm thick as prepared, base and acid-treated 1T' MoS$_2$ membranes. Permeate conductivity in the case of a 2H MoS$_2$ membrane is also shown. (b) Reversible change of Na$^+$ permeation rate through a MoS$_2$ membrane treated with acid and base alternatively. The Na$^+$ concentration at the permeate side was determined by ICP-MS.

In addition to ion diffusion experiments, we also performed pressure filtration of salt solutions using acid and base treated 1T' MoS$_2$ membranes (Supplementary Fig. 9). The membrane filtration performance is evulated by calculating the salt rejection. The salt rejection was calculated as (1 − $C_P/C_F$), where $C_p$ is the salt concentration at the permeate side and $C_F$ is the salt concentration at the feed side. The amount of salts permeated were measured by probing the concentration of salt in the permeate side by ICP-MS (for MgCl$_2$ and trisodium citrate) or by ultraviolet–visible (UV–vis) absorption (for K$_3$[Fe(CN)$_6$] and Na$_4$PTS) of the permeate water. Taking potassium ferricyanide (K$_3$[Fe(CN)$_6$]) as an example, MoS$_2$ membrane treated with 0.1 M LiOH shows an 18% rejection of K$_3$[Fe(CN)$_6$]. The rejection of K$_3$[Fe(CN)$_6$] suddenly increased to 68% after treating a MoS$_2$ membrane at pH 3. Further treating the membrane with pH 2 solution led to 97% rejection of K$_3$[Fe(CN)$_6$]. As expected, rejection also follows a hysteresis depending on the direction of pH treatment, and it follows the reverse trend of water flux through the membrane (Supplementary Fig. 9a). Other salt solutions, such as MgCl$_2$, and trisodium citrate, show a similar trend (Supplementary Fig. 9b). However, the rejection of pyrenetetrasulfonic acid tetrasodium salt (Na$_4$PTS) remained high throughout the pH span due to the larger hydrated diameter (comparable to the interlayer channel. ~ 1 nm) of the PTS ion.

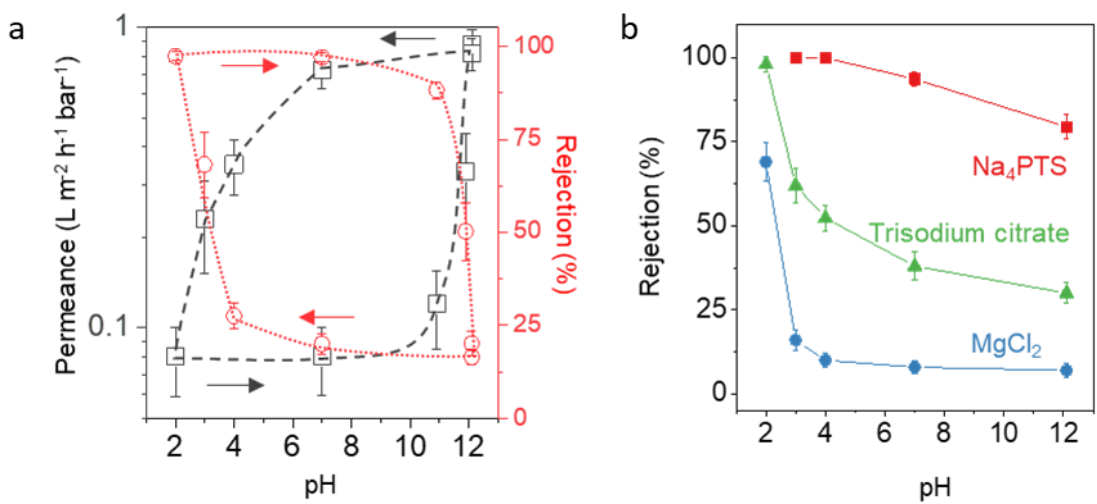

**Supplementary Fig. 9. pH-controlled salt rejection through a 1T' MoS$_2$ membrane.** (**a**) Water flux (in black) and rejection (in red) of 20 mM K$_3$[Fe(CN)$_6$] as a function of pH treatment through a 500 nm thick 1T' MoS$_2$ membrane (1 cm$^2$ area) under 5 bar pressure in a dead-end permeation. The arrows denote the pH change direction. (**b**) Rejection of 20 mM MgCl$_2$, trisodium citrate and pyrenetetrasulfonic acid tetrasodium salt (Na$_4$PTS) at 5 bar pressure through 1T' MoS$_2$ membrane (area of 1 cm$^2$ and thickness of 0.5 µm) as a function of pH treatment. Error bars in **a** and **b** denote standard deviations using three different samples.

8. Application of MoS$_2$ membrane for autonomous wound monitoring

Wound healing is a complicated problem that requires constant monitoring. A wound is typically monitored by visual inspection during dressing change, which can disturb the wound healing process causing further injury to the wound. Hence, a novel autonomous way to monitor wounds have a drastic impact on wound management and healthcare cost[3]. Monitoring the pH of a wound has been the focus of several recent studies[4-7] since the change in the pH of the wound site can provide useful information regarding the status of the wound. For example, a healing wound approaches an average pH of 4.7, similar to the pH recorded for the skin[8]. In the case of an infected wound, the pH rises above 7, with chronic wounds reaching a pH of 9[9-11]. Here, we used the pH-responsive permeation properties of 1T' MoS$_2$ membrane to build a pH sensor for wound monitoring. As a proof-of-concept device, MoS$_2$ membranes were used to detect pH of a simulated body fluid (SBF).

The devices were fabricated by sandwiching a 1T' MoS$_2$ membrane between a transparent acrylic plate and polyester sheet (PE) sheet with tape adhesives, at the edges on which the pH indicator paper was carefully mounted as depicted in the schematic Supplementary Fig. 10a. The adhesive tapes were stacked such that they kept the pH paper in close contact with the MoS$_2$ sample. The top acrylic plate had a rectangular access hole carved at the centre, which acted as a reservoir to hold up to 20 µL of the solutions tested. The air gap between the MoS$_2$ sample and the bottom of the acrylic cover was sealed with a thin layer of vacuum grease to prevent the water from flowing laterally. A photograph of a completed device is shown in the inset of Supplementary Fig. 10a. The pH of SBF solutions was adjusted to simulate different stages of an infected wound in the healing process. SBF adjusted at pH 4-4.3 corresponds to a healing wound, whereas SBF adjusted at pH 7 and 9 corresponds to the onset and peak of chronic wound status, respectively. A 0.1 M HCl and 0.1 M NaOH solutions measuring pH 1 and 12.1 were used as reference points for this study. Both acid-treated (black outline in Supplementary Fig. 10b) and as-prepared/base treated (red outline in Supplementary Fig. 10b) MoS$_2$ membranes were tested for this purpose (Supplementary Fig. 10b). We also tested the permeation of pH 2 and pH 4 SBF solution through a GO membrane as a control (blue outline). For the base treated membranes, permeation and hence the respective colour change indicative of the pH was observed for the SBF solutions at pH 7.0, 9.0 and the reference solution 0.1M NaOH at pH 12.1. Neither pH 4 SBF nor 0.1M HCl permeates through these MoS$_2$ membranes. On the other hand, only 0.1M NaOH solution permeated through acid treated MoS$_2$ membranes. This observation mirrored our water vapour permeation results. The spontaneous permeation switching for the base treated sample at pH 7 and above will be very valuable for use in autonomous infection sensing in wounds since the membrane only responds if the wound is infected (pH above 7). At normal wound conditions, the membrane will be impermeable to wound exudate. As a control experiment, we used GO coated PES membranes for pH sensing and found that acidic SBF solutions with pH 2 and 4 permeate and, hence, are unsuitable for autonomous wound sensing applications. Further work is needed to integrate MoS$_2$ membranes or coating with wound dressing materials to detect the wound pH *in situ.*

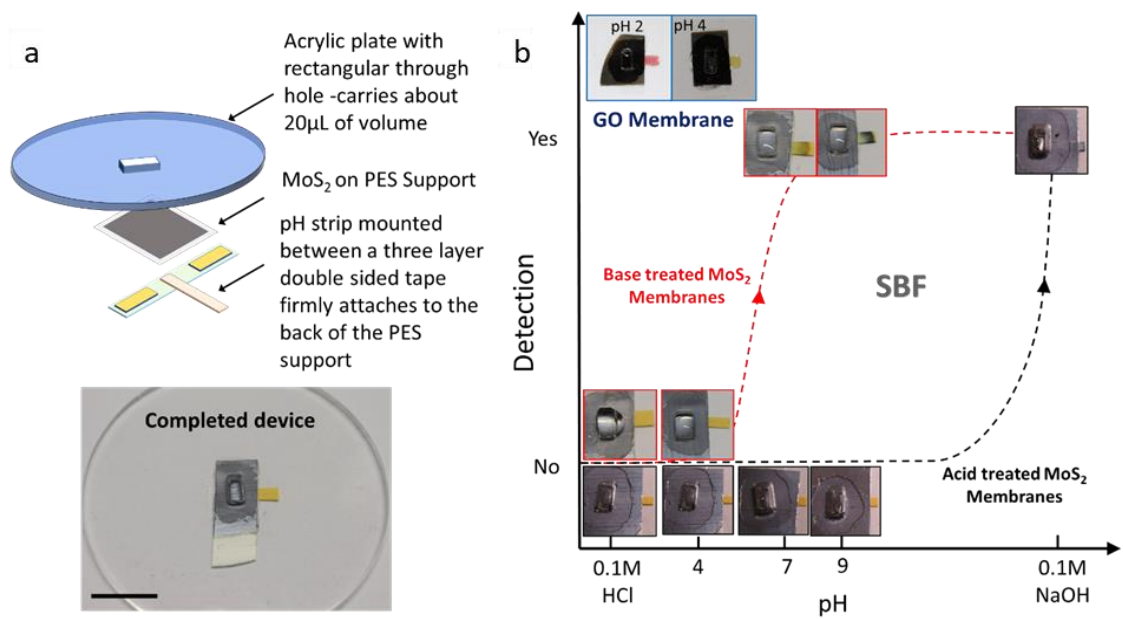

**Supplementary Fig. 10. Application of pH responsive MoS$_2$ membranes for wound infection detection.** (**a**) Schematic representation of the device fabricated for proof-of-concept studies of pH detection with simulated body fluid (SBF). Bottom inset: Photograph of a completed device with MoS$_2$ membrane and pH paper for indication. Scale bar- 1 cm. (**b**) Photograph of the fabricated devices with pH adjusted SBF solution. The pH of the corresponding solution is shown on the X-axis. Both base-treated (red outline photographs) and acid-treated (black outline photographs) MoS$_2$ membranes were employed. The colour change on the pH paper of base treated samples indicates SBF permeation and its absence in acid treated samples indicates absence of SBF permeation. Red and black dotted lines are guides to the eye. Top inset shows permeation of pH 2 and pH 4 SBF solutions through GO membranes (blue outline photographs).

The ability of 1T' MoS$_2$ membranes to respond to the pH corresponding to the infected wound in an autonomous way is a substantial advantage for smart drug delivery applications since it can be used to trigger a treatment (e.g., releasing an antibiotic agent from the membrane) according to the physiological pH of the wound, i.e. only when there are natural signs of infection. Further work is needed to demonstrate this application.

9. X-ray diffraction of 1T' MoS$_2$ membranes exposed to different pH treatment

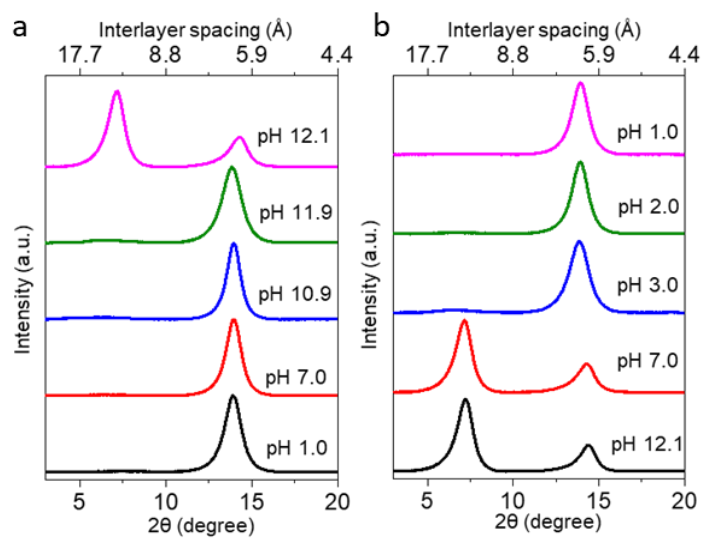

**Supplementary Fig. 11. X-ray diffraction of 1T' MoS$_2$ membranes exposed to different pH treatment.** X-ray diffraction of 1T' MoS$_2$ membranes treated with aqueous solutions with pH changing from 1 to 12.1 (**a**) and 12.1 to 1 (**b**).

10. XPS of acid and base treated 1T' MoS$_2$ membranes

Supplementary Fig. 12 shows the Mo 3d core level XPS spectra of acid and base treated 1T' MoS$_2$ membranes. These spectra were fitted as described previously in section 1. The phase fraction analysis on both the acid and base treated samples as determined by deconvoluting the Mo 3d peak (Supplementary Fig. 12) indicates the absence of any significant phase changes due to acid or base treatment. Compared to the as-prepared 1T' MoS$_2$ membranes, the acid-treated MoS$_2$ membrane only shows a slight decrease in the 1T' proportion from 56% to 46%, but it recovered to 56% after the base treatment (Supplementary Fig. 12). Additionally, XPS characterization also rules out the possibility of oxidation of MoS$_2$ during acid or base treatment since no signatures of MoO$_x$ species were found in the XPS spectrum.

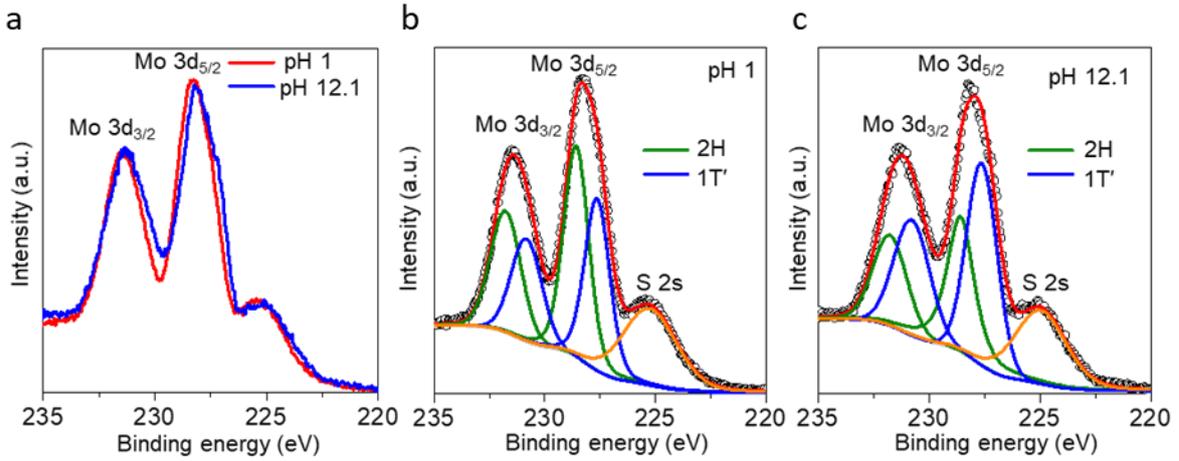

**Supplementary Fig. 12. X-ray photoelectron spectroscopy of acid and base treated 1T' MoS$_2$ membranes.** (**a**) Raw XPS spectra at the Mo 3d core level of the acid (pH 1) and base (pH 12.1) treated 1T' MoS$_2$ membranes. (**b**) XPS peak fitting for acid and (**c**) base treated 1T' MoS$_2$ membranes. Black circles- raw data; red line- the fitting envelope; blue line and green lines, deconvolved peaks attributed to the 1T' and 2H phases respectively; orange line denotes the S 2s peak fitting.

11. X-ray absorption spectroscopy (XAS) measurements

The XAS at the Mo K-edge were recorded at beam line BL14W1 of the Shanghai Synchrotron Radiation Facility (SSRF), China. The electron storage ring was operated at 3.5 GeV. 1T', 2H, acid and base treated 1T' MoS$_2$ membranes were directly used for measurement in the transmission mode using a Si (311) double-crystal monochromator in ambient conditions. Mo foil was used as a reference for energy calibration. The beam size was limited by the horizontal and vertical slits with an area of 1 × 4 mm$^2$ during XAS measurements. The Mo K-edge spectra were recorded from 19800 to 20800 eV with a total scan time of ca. 30 min.

Data analysis was performed according to established methods[12] using WinXAS version 3.1. Edge data were normalized by fitting the smoothly varying parts of the absorption spectrum both below and above the edge to tabulated X-ray cross-sections, using a single polynomial background and scale factor. The X-ray absorption fine-structure (EXAFS) data were converted to k space using an initial E$_0$ value of 20000 eV for Mo. The k space data were weighted by k$^2$ and Fourier transformed to R-space with the k-space ranging from 2.5 to 12.5 Å$^{-1}$.

Simulations of EXAFS scattering paths to conduct fits of the experimental data were calculated using the *ab-initio* self-consistent field computational code Feff 8 program[13] according to the following EXAFS equation[12]:

$$\chi(\mathbf{k}) = \sum_j \frac{S_0^2 N_j}{k R_j^2} \times e^{-\frac{2R_j}{\lambda_j(k)}} \times e^{-2\sigma_j^2 k^2} \times f_j(k) \times \sin[2kR_j + \delta_j(k)]$$

where f(k) and δ(k) are scattering properties of the neighbouring atoms. f(k): backscattering amplitude; δ(k): phase shift. k is defined as $\mathbf{k} = \sqrt{2m(E-E_0)/\hbar^2}$. N is the number of neighboring atoms (coordination number). R is the distance to the neighboring atom (bond distance). σ$^2$ is the disorder in the neighbor distance (Debye Waller factor). S$_0^2$ and λ are reduction factor and mean free path. j accounts for different paths of scattering[14,15]. Uncertainties in EXAFS fitting parameters were computed from off-diagonal elements of the correlation matrix and weighted by the square root of the reduced chi-squared value obtained for the simulated fit. The amount of experimental noise from 15-25 Å in R-space was also taken into consideration for each EXAFS spectrum[16].

Supplementary Fig. 13 shows Mo K-edge X-ray absorption near edge structure spectra, extended X-ray absorption fine-structure (EXAFS) spectra, and Fourier transformed EXAFS spectra of $MoS_2$ membranes. Fourier transformed EXAFS spectra of 2H $MoS_2$ has mainly two peaks corresponding to a Mo-Mo bonding length of 3.17 Å, and a Mo-S bond length of 2.41 Å. The spectrum from the as-prepared 1T' $MoS_2$ membrane shows two Mo-Mo bond distances of 2.8 and 3.17 Å and one Mo-S bond length of 2.41Å. The presence of two peaks corresponding to two different Mo-Mo bond lengths in contrast to a single Mo-Mo bond distance in the 2H phase confirmed that the as-prepared $MoS_2$ membranes primarily consist of the 1T' phase[17,18]. Even though the EXAFS spectrum obtained from the base treated sample resembles to that of pristine 1T' $MoS_2$ membrane, after acid treatment the spectrum shows significant changes in the two Mo-Mo bond contributions. Further EXAFS curve fitting analysis suggests that these changes are due to the change in Mo co-ordination numbers. Supplementary Table 1 summarizes the curve fitting results for the EXAFS spectra. The co-ordination number obtained from the curve fitting shows significant changes in each bond after the acid treatment, suggesting local atomic structural changes in the $MoS_2$.

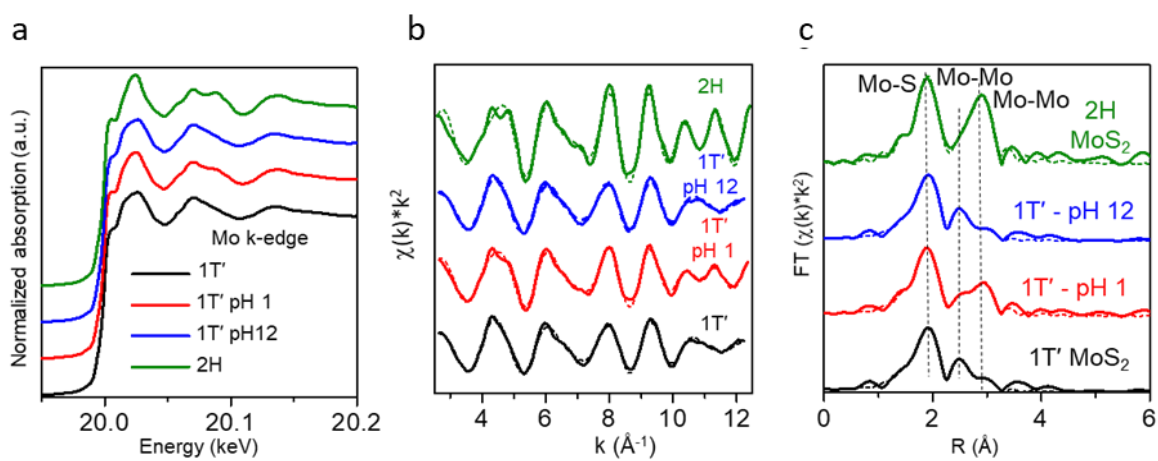

**Supplementary Fig. 13. X-ray absorption spectroscopy (XAS).** (**a**) Mo K-edge X-ray absorption near edge structure (XANES) spectra, (**b**) extended X-ray absorption fine-structure (EXAFS) spectra, and (**c**) Fourier transformed EXAFS spectra of 1T', 2H, acid, and base treated 1T' $MoS_2$ membranes (colour coded labels). The dotted lines show the data fitting. The quantified fitting results are shown in Supplementary Table 1. Fourier transformed EXAFS spectra represent radial distribution functions around the Mo atoms in the $MoS_2$ membranes. The peak positions shown by the dashed lines correspond to interatomic distances. The peak positions are slightly shifted from the actual interatomic distances because of the phase shift effects. All these changes in the EXAFS spectrum after acid treatment were reversible, and the structure recovered to that of pristine 1T' $MoS_2$ after base treatment.

**Supplementary Table 1.** Mo K-edge EXAFS curve fitting results of 1T', 2H, acid and base treated 1T' $MoS_2$ membranes.

| Sample | Shell | CN | R / Å | $\Delta\sigma^2$ / x$10^{-3}$Å$^2$ | $\Delta E_0$ / eV |
|---|---|---|---|---|---|
| 1T' $MoS_2$ | Mo-S | 5.5±0.7 | 2.415±0.007 | 5.2±1.1 | -3.7±0.9 |
| | Mo-Mo | 2.7±0.7 | 2.805±0.006 | 6.9±1.7 | 2.2±1.4 |
| | Mo-Mo | 3.0±1.2 | 3.169±0.007 | 7.5±2.0 | 2.8±1.4 |
| 1T' $MoS_2$ pH 1 | Mo-S | 5.5±0.7 | 2.409±0.009 | 4.5±1.3 | -2.5±1.5 |
| | Mo-Mo | 0.6±1.4 | 2.808±0.008 | 5.0±1.3 | 0.4±2.7 |
| | Mo-Mo | 4.1±1.7 | 3.170±0.007 | 4.3±2.9 | 3.7±1.1 |
| 1T' $MoS_2$ pH 12 | Mo-S | 5.7±0.7 | 2.418±0.007 | 5.4±1.1 | -3.5±0.9 |
| | Mo-Mo | 2.6±0.7 | 2.803±0.006 | 6.5±1.7 | 2.2±1.4 |
| | Mo-Mo | 2.8±1.2 | 3.170±0.007 | 7.2±2.0 | 2.5±1.4 |
| 2H $MoS_2$ | Mo-S | 6.0 | 2.414±0.007 | 2.9±0.9 | -0.3±0.9 |
| | Mo-Mo | 6.0 | 3.169±0.006 | 2.1±0.5 | -0.5±1.2 |

CN, coordination number; R, bonding distance; $\Delta\sigma^2$, Debey-Waller factor; $\Delta E_0$, inner potential shift, amplitude reduction factor $S_0^2$ was set as 0.83 for all the samples.

In order to further understand the reason for the changes in the co-ordination number and structural changes of $MoS_2$ after the acid treatment, we simulated EXAFS spectra using the known structural models of 2H, 1T, 1T' and 1T'' $MoS_2$ on the Feff 8 program[13]. Supplementary Fig. 14 shows different structural models of different phases of $MoS_2$ and the corresponding simulated EXAFS spectra. These show that both 1T and 2H $MoS_2$ has a Mo-Mo bond length of 3.17 Å whereas 1T' $MoS_2$ has a Mo-Mo bond length of 2.8 Å, consistent with the experimental data. However, in 1T'' $MoS_2$ there exist two types of Mo atoms with different co-ordination numbers[18]. As shown in supplementary Fig. 14d, Mo atoms with teal color have the same co-ordination environment with that of Mo in 1T' $MoS_2$ whereas Mo atoms with orange color do not have a 2.8 Å Mo-Mo bond. Thus, the overall co-ordination number of 2.8 Å Mo-Mo bonds decreased compared with the corresponding co-ordination number in 1T' $MoS_2$. This suggests that the decrease in the co-ordination number of 2.8 Å Mo-Mo bond after the acid treatment is due to the partial conversion of 1T' structure into 1T''. Consistent with this, the simulated EXAFS spectra from the 1T'' $MoS_2$ agrees well with the spectra from the acid treated sample.

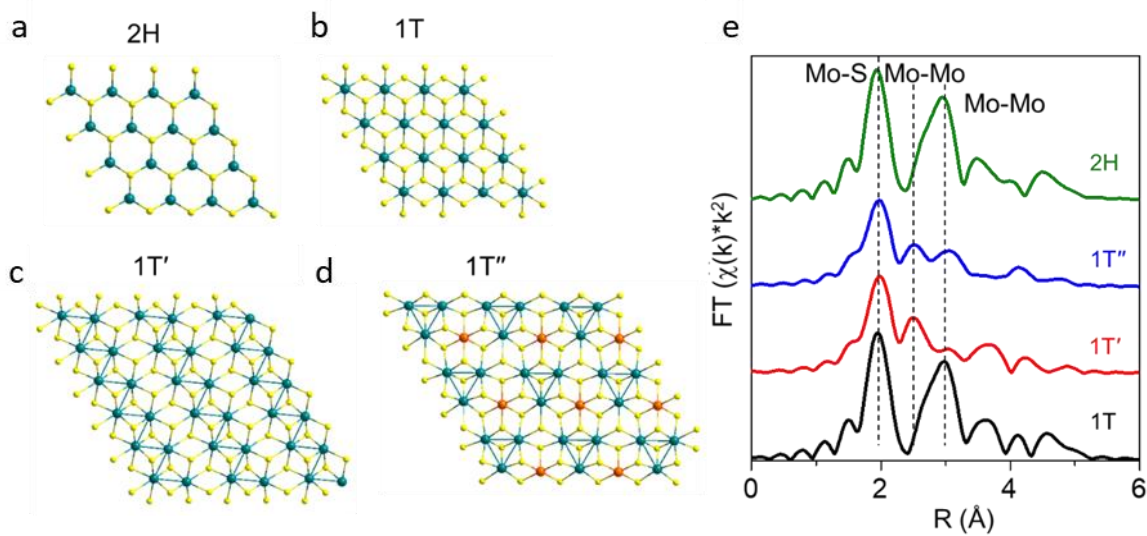

**Supplementary Fig. 14. EXAFS simulation.** Structure models of (**a**) 2H, (**b**) 1T, (**c**) 1T' and (**d**) 1T'' $MoS_2$. Colour code: teal, Mo; yellow, S; (orange color in 1T'' represent Mo with coordination environment different from Mo in teal color). (**e**) Simulated EXAFS spectra using the structure models in **a**-**d**.

12. Influence of different cations

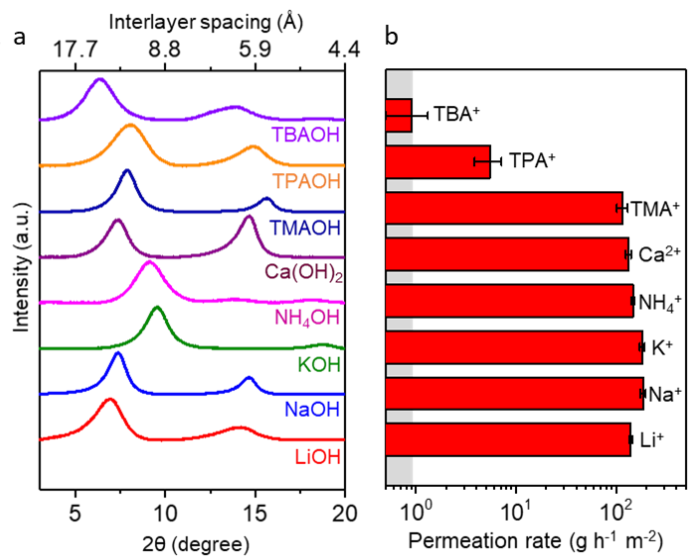

**Supplementary Fig. 15. Influence of different cations on the interlayer spacing and water permeation.** (**a**) X-ray diffraction and (**b**) water permeation rate of 1T' $MoS_2$ membranes treated with 0.1 M concentration base solution with different cations. Gray area indicates our detection limit. Error bars denote standard deviations using three different samples.

To further investigate the effect of other cations on water permeation through $MoS_2$ membranes, we have also performed XRD and water permeation on 1T' $MoS_2$ membranes treated with alkaline solutions prepared from different bases, e.g. LiOH, NaOH, KOH, $Ca(OH)_2$, $NH_4(OH)$, tetramethylammonium hydroxide (TMAOH), tetrapropylammonium hydroxide (TPAOH), and tetrabutylammonium hydroxide (TBAOH) (Supplementary Fig. 15). All the base treatments were performed with 0.1 M base solutions after soaking the as-prepared 1T' $MoS_2$ membranes in 0.1M HCl to remove intercalated $Li^+$ ions. Supplementary Fig. 15b shows that the water permeation rate through MoS$_2$ membranes treated with different cations is mostly independent of the cations even though they have a slightly different interlayer spacing (Supplementary Fig. 15a) due to the varying sizes of the cations[19]. The only exception for this was the weakly hydrating cations such as TBA, where the water permeation rate was low in spite of the membrane having a larger interlayer spacing of 13.9 Å. This could be due to the hydrophobic nature of TBA compared to other cations[20].

In addition to water permeation and XRD, we have also studied adsorption of other cations (Na$^+$ and K$^+$) on 1T' MoS$_2$ membranes as a function of pH treatment by using inductively coupled plasma mass spectrometry (ICP-MS). Our experiments confirm that similar to the case of Li$^+$ ions, Na$^+$ and K$^+$ cations also get adsorbed into the MoS$_2$ membranes and the adsorption shows a pH-responsive hysteresis (Supplementary Fig. 16), proving that pH dependent ion adsorption, and hence water permeation is not exclusive to Li$^+$ ions.

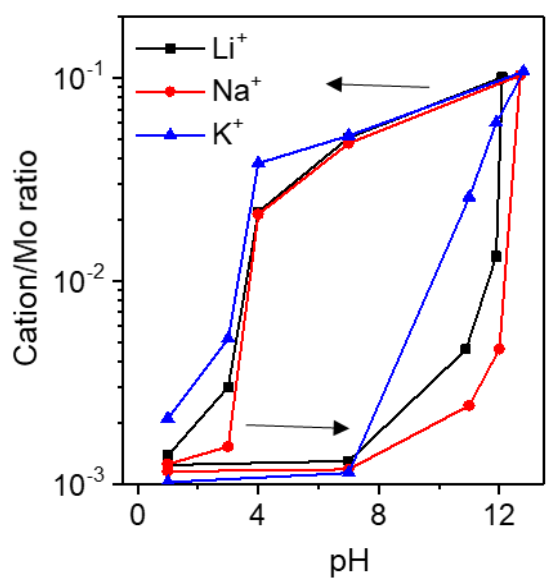

**Supplementary Fig. 16. pH-dependent cation adsorption on MoS$_2$ membranes.** Cation/Mo ratios of MoS$_2$ membranes treated with LiOH, NaOH and KOH aqueous solutions under different pH determined by ICP-MS. Arrows indicate the direction of pH change.

13. <u>Water vapour adsorption measurements</u>

Water vapour adsorption isotherms for 1T' MoS$_2$ membranes were measured in a Hiden IGA gravimetric analyser at 20 °C. In a typical experiment, small pieces of MoS$_2$ membranes (5 mg) were exposed to different humidity at 20 °C, allowing sample mass to reach gravimetric equilibrium at each step. The equilibrium amount of vapor adsorbed per unit mass is plotted as a function of the relative vapour pressure (p/p$_0$) (Supplementary Fig. 17a), at constant temperature and pressure to generate an isotherm. Here p$_0$ represents the saturation vapour pressure of water at 20 °C.

Supplementary Fig. 17a shows the water vapour adsorption isotherms of MoS$_2$ membranes treated at different pH. The adsorption isotherm of the 1T' MoS$_2$ treated with base is of type II, which is typical for monolayer-multilayer adsorption in non-porous solids. The isotherm consists of two distinct steps. At low humidity, water molecules get adsorbed on to the surface forming multilayers of water inside the capillary. Then, at higher humidity, there occurs a cooperative process involving the interaction between adsorbate molecules which leads to capillary condensation and thus sudden increase in water uptake. Whereas acid-treated MoS$_2$ membranes show negligible water uptake even at a p/p$_0$ of 0.99. Supplementary Fig. 17b illustrates the water uptake at p/p$_0$ of 0.99 as a function of the pH treatment.

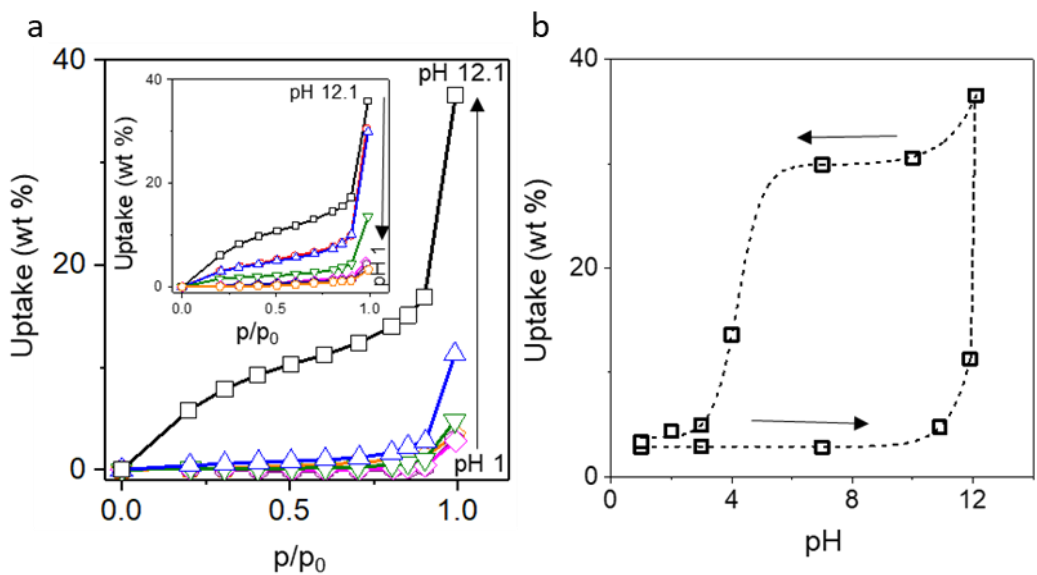

**Supplementary Fig. 17. pH-dependent water vapour adsorption.** (**a**) Water vapour adsorption isotherm of MoS$_2$ membranes treated at different pH from 1 to 12.1. p/p$_0$ is the ratio of vapour pressure to saturated vapour pressure at a fixed temperature. Inset: water vapour adsorption isotherm of MoS$_2$ membranes treated at different pH from 12.1 to 1. (**b**) Water uptake at p/p$_0$ = 0.99 as a function of pH. Dashed line is a guide to the eye and arrows indicate the direction of pH change.

14. Role of base in the lithiation process

To confirm whether the base is important for the lithiation process, 2 µm thick 1T' MoS$_2$ membranes were first treated with 0.1M HCl followed by 0.1 M LiCl and 0.1 M LiOH solution successively. X-ray diffraction (XRD) studies and water vapour permeation through these membranes were checked. As expected, membranes treated with 0.1 M HCl showed disappearance of low angle XRD peak and inhibition of water molecules. Subsequent LiCl and LiOH treatment of these membranes resulted in completely different properties. Whereas LiCl treated membranes showed no difference in XRD or permeation properties from the HCl treated membrane, LiOH treated membranes recovered their low angle XRD peak and water permeation, suggesting basic pH is required for the change in permeation properties.

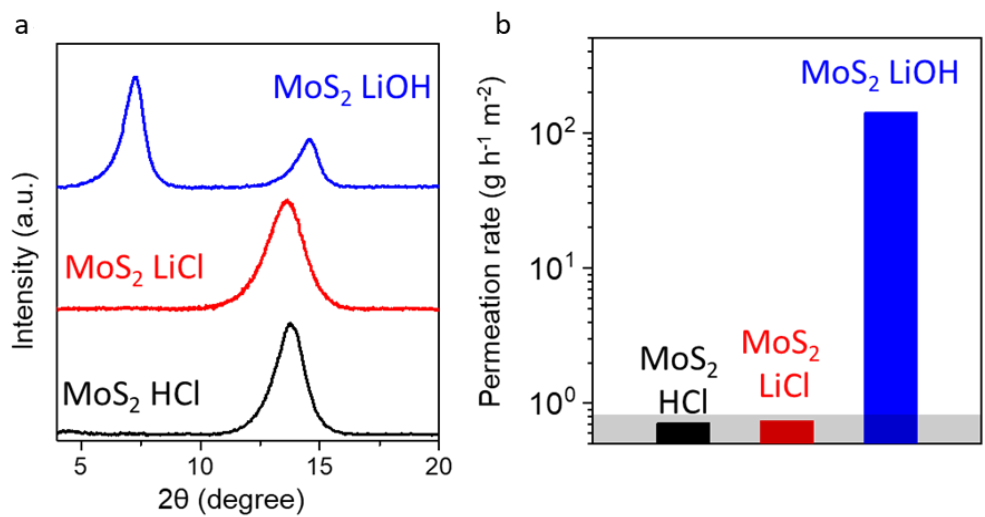

**Supplementary Fig. 18. Role of base in the lithiation process.** (**a**) XRD and (**b**) water permeation rate of a MoS$_2$ membrane treated with HCl, LiCl and LiOH solutions successively. The membrane remained impermeable when treating the acid-treated membrane with LiCl solution, confirming the role of base in the lithiation process. Gray shaded area is the detection limit.

15. Activation behaviour of Li$^+$ ion adsorption

To gain an insight into the protonation/delithiation or deprotonation/lithiation process, we treated 1T' MoS$_2$ membranes with 0.1 M LiOH at different temperatures, and then studied the Li/Mo ratios using ICP-MS. Supplementary Fig. 19 shows that the Li$^+$ ion adsorption decreases exponentially with increasing temperature and follows an Arrhenius behaviour with an energy barrier of 30 kJ/mol.

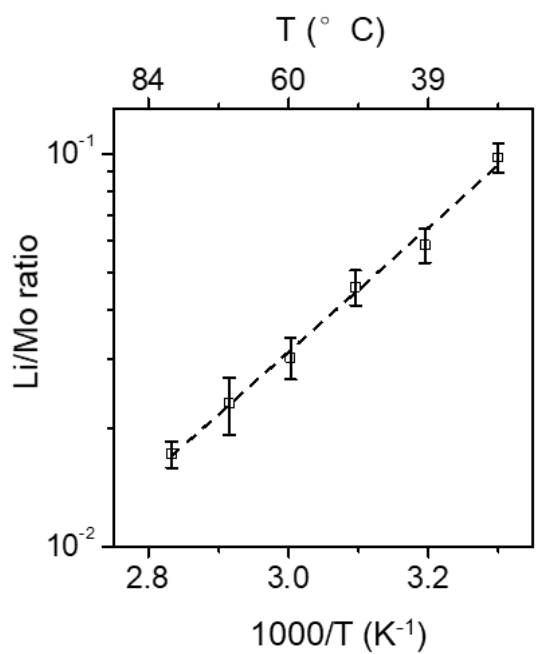

**Supplementary Fig. 19. Temperature dependence of cation adsorption.** Li/Mo ratio of a 1T' MoS$_2$ membrane after treating with 0.1 M LiOH at different temperatures. Dashed line is the best fit to the Arrhenius behavior. Error bars denote standard deviations using three different samples.

16. Hydrogen plasma treatment of MoS$_2$ membranes

In order to validate that the pH-responsive permeation of water through 1T' MoS$_2$ membranes occurs via protonation and deprotonation of sulphur atoms, we exposed the membrane surface to a hydrogen plasma under 50 mTorr pressure at 100 W for 30 min in a Cobra PlasmaPro RIE instrument by Oxford Systems. As expected, the hydrogenated membrane showed a complete blockage of water transport (similar to acid-treated 1T' MoS$_2$ membranes) (Supplementary Fig. 20). The permeability of the membrane returned to its as-prepared value after exposing the membrane to 0.1 M LiOH solution for 1h.

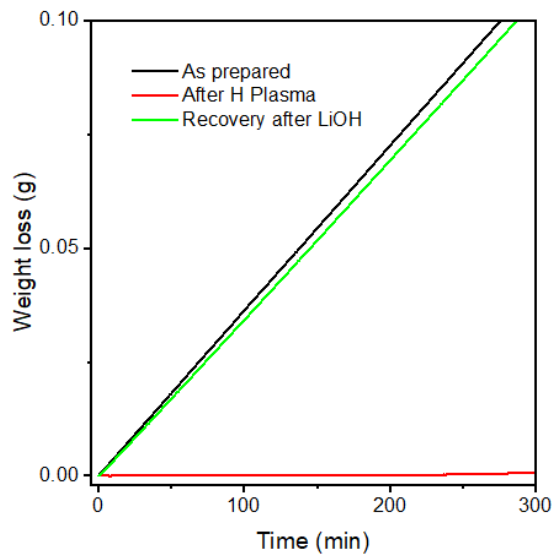

**Supplementary Fig. 20. Effect of hydrogen plasma treatment.** Weight loss for a water filled container sealed with a 2 µm thick 1T' MoS$_2$ membrane before and after exposure to hydrogen plasma. The permeation recovered after exposing the hydrogen plasma treated sample to LiOH solution.

17. Ethanol vapour adsorption and permeation measurements

Ethanol adsorption isotherms for 1T' MoS$_2$ membranes were measured in a Hiden IGA gravimetric analyser at 20 °C as described in Section 13. Supplementary Fig. 21a shows the ethanol vapour adsorption as a function of concentration of the ethanol. At p/p$_0$ of 0.99, a sudden increase in ethanol uptake (~ 19%) is noticed, signifying the condensation of ethanol in the interlayer spaces of the MoS$_2$ membrane. In agreement with this, our XRD analysis also confirms that ethanol gets intercalated between MoS$_2$ layers. Supplementary Fig. 21b shows the XRD of MoS$_2$ membrane before and after the exposure to ethanol. The shift of the low angle peak from 7.7° to 6.1°, corresponding to an increase in interlayer spacing to 14.4 Å from 11.5 Å, suggests the intercalation of ethanol between MoS$_2$ layers. Even though ethanol gets intercalated and adsorbed similar to water, surprisingly its permeation rate was two orders of magnitude lower than that of water (Supplementary Fig. 22).

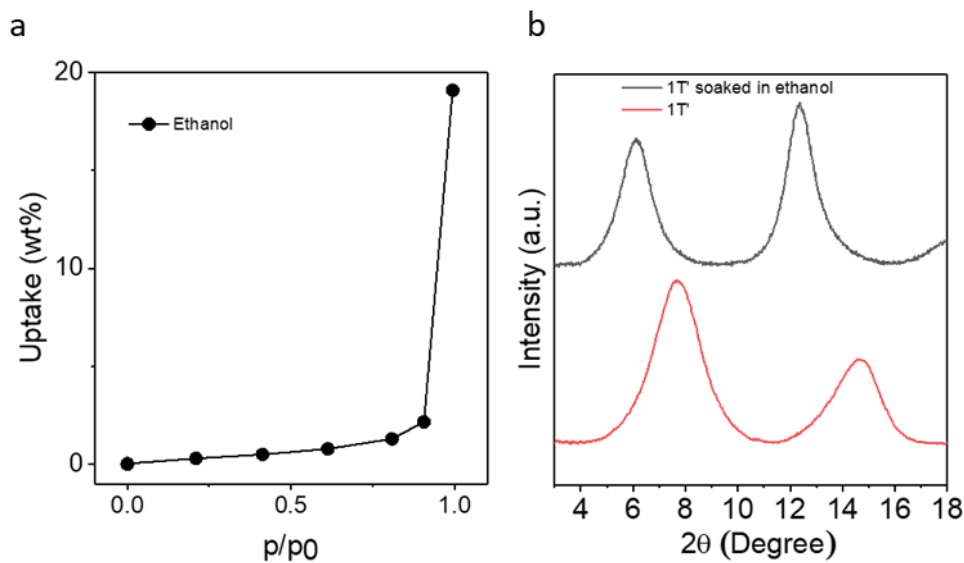

**Supplementary Fig. 21. Ethanol adsorption and intercalation into 1T' MoS$_2$ membranes. (a)** Ethanol adsorption isotherm of a 1T' MoS$_2$ membrane at 20 °C. (**b**) X-ray diffraction pattern of an as-prepared 1T' MoS$_2$ membrane before and after soaking in ethanol.

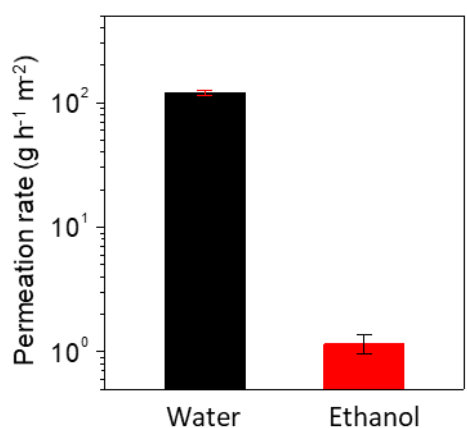

**Supplementary Fig. 22. Ethanol permeation through 1T' MoS$_2$ membranes.** Comparison of ethanol and water permeation rate through 1T' MoS$_2$ membranes. Error bar is the standard deviation from three different measurements.

18. <u>Ab Initio Molecular Dynamics (AIMD) Simulations</u>

Ab initio MD simulations of water confined between MoS$_2$ were performed using the CP2K[21] software package. Simulations typically contained ~500 atoms, from 1-16 Li atoms and between 103-108 water molecules. The optB88-vdW[22,23] functional was used, with Goedecker-Teter-Hutter pseudopotentials[24], a 550 Ry plane-wave cut-off, a relative cut-off of 60 Ry and a DZVP molecularly optimised basis set[25]. Prior to running AIMD simulations, water-MoS$_2$ systems were annealed using classical molecular dynamics in the LAMMPS software package[26], using the TIP4P/2005 water model[27] and a Lennard-Jones potential for the water-lithium interactions[28]. Equilibration using AIMD was performed in the NVT ensemble, using a chain of 5 Nosé-Hoover thermostats to maintain a temperature of 500 K for 5 ps. This was followed by a second equilibration stage at 400 K for a further 5 ps. Final data collection for analysis was performed at 400 K over a 20 ps period.

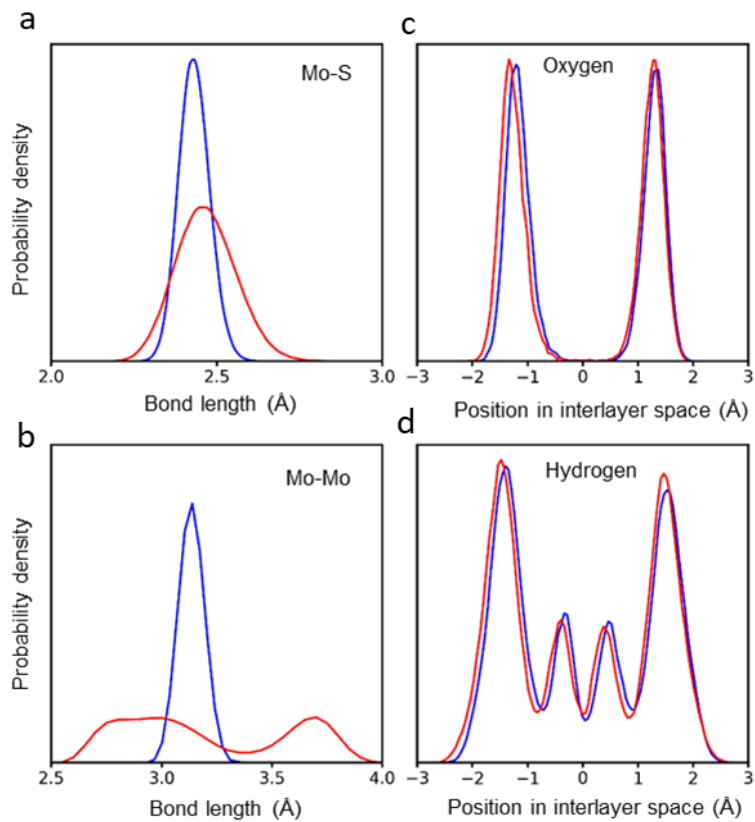

**Supplementary Fig. 23. Water between 2H and 1T' MoS$_2$ laminates.** Mo-S (**a**) and Mo-Mo (**b**) bond lengths for 2H (blue) and 1T' phases (red) of MoS$_2$. The very different stiffnesses of the Mo-S bonds in the two polymorphs give rise to peaks of very different broadness. The rigid geometry of the 2H polymorph gives rise to a single Mo-Mo separation while the thermal disorder and continuous reorganisation of Mo atoms in the 1T' phase give rise to a broad bimodal distribution. Comparison of water oxygen probability distributions (**c**) and water hydrogen probability distributions (**d**) for 2H (blue) and 1T' phases (red) showing almost identical bilayer arrangements.

Supplementary Fig. 23 shows the influence of two different polytypes (2H and 1T') on the structure of water confined between them. As can be seen (Supplementary Figs. 23a, b) the two polymorphs have very different characteristic bond lengths and bond length distributions. Despite this, the structure of the intercalated water is almost identical in both cases (Supplementary Figs. 23c, d).

To probe the influence of Li in between MoS$_2$ layers on the structure of water and MoS$_2$, we performed simulations with Li/Mo ratios ranging from 0.02-0.3. Our results typically indicate the minimal variation between the structure of the interstitial water for a range of low lithium concentrations studied. We, therefore, choose to give results only for the low Li concentration (Li/Mo=0.02) and high Li concentration (Li/Mo = 0.3) cases, for which we see a dramatic difference in the structuring. In both cases, water forms a clearly ordered bilayer structure, with locally tetrahedral arrangements around solvated lithium atoms. The primary difference between the results of the two simulations lies in the position of Li ions, orientation of the water molecules in these bilayers, and the structure of MoS$_2$ layers itself. Supplementary Fig. 24 shows that the Li atoms were found to localise between the two water monolayers, as compared to at low concentration (small Li/Mo ratio) when they are found to localise close to the MoS$_2$ surface. Supplementary Fig. 25 shows the effect of changing lithium concentration on both the structure of the intercalated water and the MoS$_2$ itself. As discussed in the main text, upon increasing the concentration of Li, a phase change in the intercalated water bilayer is observed (Supplementary Figs. 25c, d) manifested as a rotation of the water molecules such that hydrogen atoms point towards the MoS$_2$ surface. Simultaneously, we can see that an increase in the ordering of the MoS$_2$ substrate is observed (Supplementary Figs. 25a, b). At low lithium concentrations, the local ordering of Mo centered octahedra resembles a disordered 1T'' structure (Supplementary Fig. 14d) which constantly shifts under the influence of thermal motion, giving rise to a continuous distribution of Mo-Mo separations and a broad range of Mo-S bond lengths. In the high Li concentration case, a stable 1T' structure evolves, with well defined 'bands' of connected Mo atoms (Supplementary Fig. 14c). This effect is manifested in Supplementary Fig. 25b as a clear bimodal distribution of Mo-Mo separations, and Supplementary Fig. 25a as a narrowing of the Mo-S bond length distribution.

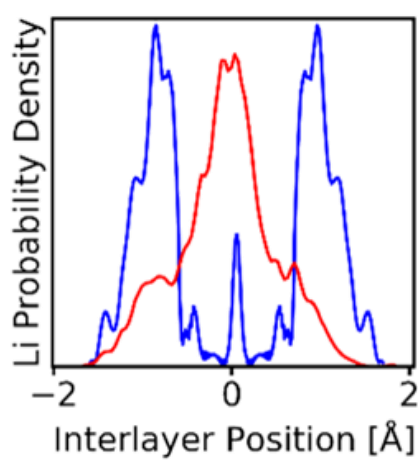

**Supplementary Fig. 24. Effect of lithium concentration on the position of Li ions between 1T' MoS$_2$ layers.** Lithium probability density in the inter-layer gallery of 1T' MoS$_2$, with the origin taken to be the center of the MoS$_2$ inter-layer space. Lithium is found to adsorb close to the MoS$_2$ surface at low Li concentrations (blue curve), while at higher concentrations (red curve) Li is found in the center of the inter-layer space.

Supplementary Fig. 26 shows the probability distributions for water oxygen atoms in the in-plane direction (that is, viewed perpendicular to the basal plane of the MoS$_2$) for two different Li concentrations. In both the low and high Li concentration cases, we observe the confined water bilayer is commensurate with the MoS$_2$ surface. Furthermore, in the high Li content case, we observed a slightly increased probability for water molecules located between Mo and S atoms suggesting an increased water hopping between different sites.

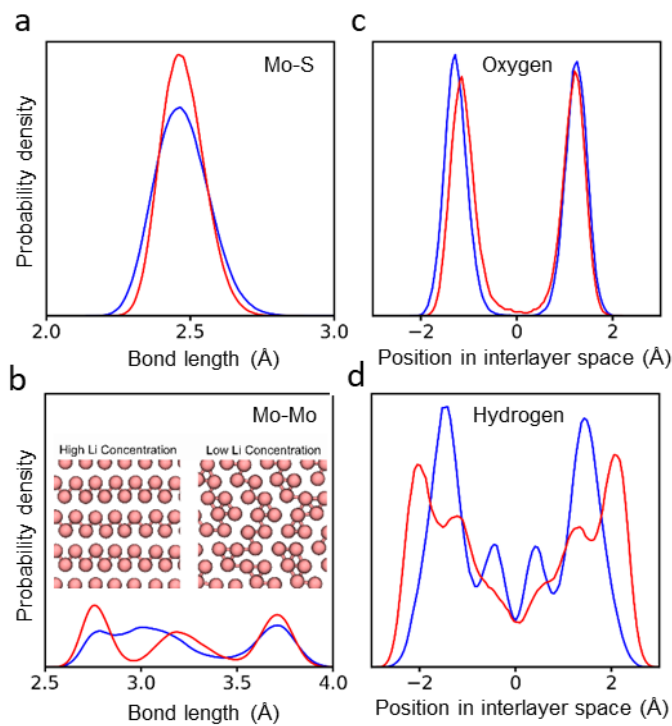

**Supplementary Fig. 25. Effect of lithium concentration on 1T' MoS$_2$ and confined water bilayer structures.** Probability density of Mo-S (**a**) and Mo-Mo (**b**) bond lengths at Li/Mo ratio of 0.02 (blue) and Li/Mo ratio of 0.3 (red). A narrowing in the distribution of Mo-S bond lengths is observed at high Li content suggesting a more ordered 1T' structure. Similarly, at low Li concentrations, the constant reordering of disordered Mo atoms causes the Mo-Mo distribution to be a broad single peak along with the Mo-Mo distribution at ~3.7 Å, while at higher Li concentrations, the distribution shows the characteristic short and long Mo-Mo bonds of the 1T' polymorph in addition to Mo-Mo distribution at ~3.7 Å. Inset: Snapshot showing ordered and disordered Mo atoms for high and low Li concentrations. The probability distribution for the position of water oxygen atoms (**c**) and water hydrogen atoms (**d**) in the 1T' MoS$_2$ interlayer space at Li/Mo ratio of 0.02 (blue) and Li/Mo ratio of 0.3 (red). The restructuring of the water bilayer involves primarily a rotation of the water molecules within the layer; thus, very little change is observed for the position of water oxygen atoms but, the position of water hydrogen atoms shows a clear reorientation, with water hydrogen atoms facing towards the MoS$_2$ substrate rather than engaging in interlayer hydrogen bonding as is observed at low Li concentration.

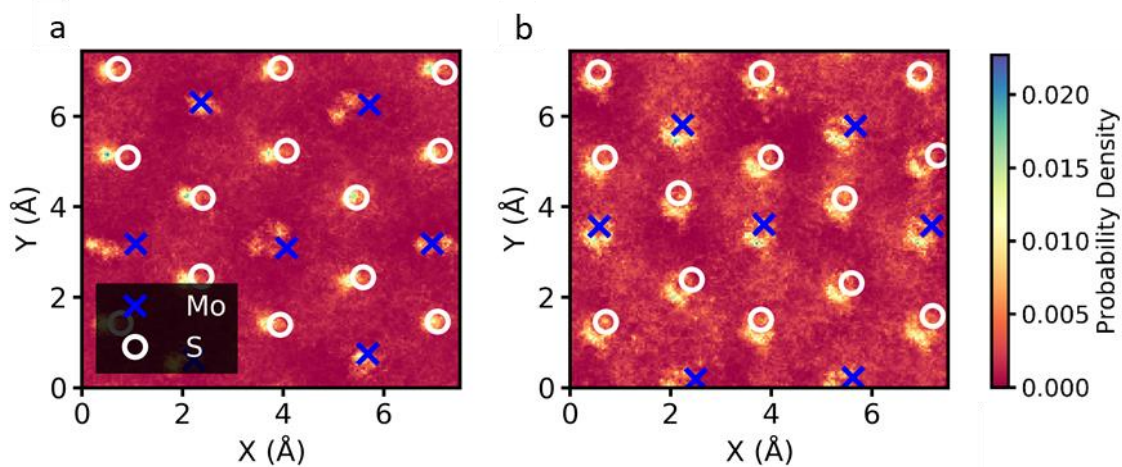

**Supplementary Fig. 26. In-plane probability distributions of water oxygen atoms on the surface of 1T' MoS$_2$ surface.** Probability distributions of water oxygen atoms on the surface of the 1T' MoS$_2$ surface for Li/Mo ratios of 0.02 **(a)** and 0.3 **(b)**. Brighter colours indicate a higher relative probability. For clarity, only a small section of the total simulated cell is shown. Average positions of MoS$_2$ molybdenum and sulphur atoms are shown as blue crosses and white circles respectively.

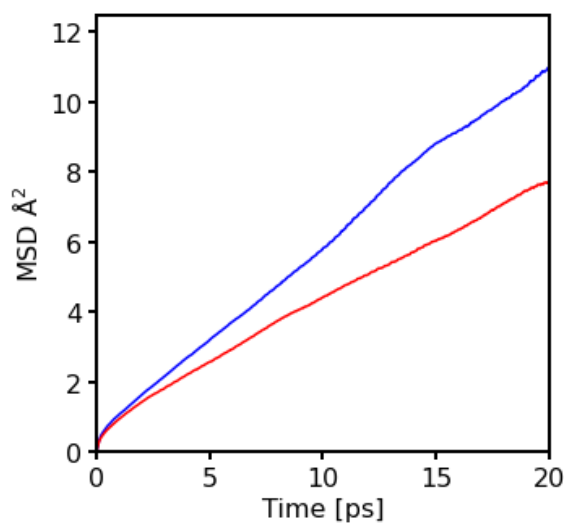

**Supplementary Fig. 27. Mean Squared Displacements for water oxygen atoms obtained from water confined within lithiated 1T' MoS$_2$.** Mean squared displacements of water oxygen atoms for low (0.02, blue) and high (0.3, red) Li/Mo ratio. There is no substantial difference in the gradient of the mean squared displacement as a function of Li/Mo ratio, indicating comparable rates of diffusion in both cases.

Supplementary Fig. 27 shows mean squared displacement (MSD) calculations for the interlayer water oxygen atoms in 1T' MoS$_2$ at low and high Li concentrations. The small observed difference in diffusion behaviour is insufficient to explain the dramatic changes in permeability observed for the macroscale membrane. This is despite the substantial change observed in the structure of the interstitial water bilayer at the two concentrations. Although it is challenging to disentangle the two effects, this observation lends credence to the hypothesis that it is the hydrostatic swelling of the MoS$_2$ membrane, induced by the presence of lithium, which affects water permeability, rather than the microscopic details of the water-water or water-MoS$_2$ interactions. This does not rule out, however, the possibility that the water structure observed in the high lithium concentration case may be a prerequisite for the thermodynamically favourable intercalation of water. The Li-induced swelling of the MoS$_2$ membrane was explored in more detail using classical molecular dynamics simulations as discussed below.

### 19. Classical molecular dynamics (MD) simulation

To understand the spontaneous intercalation of water between MoS$_2$ layers, classical molecular dynamics (MD) simulations were employed. Acid treated condition inside a 1T' MoS$_2$ laminate was mimicked by removing all the charges from the S atoms and thereby removing the Li$^+$ ions from interlayer spaces. For base treated conditions, Li$^+$ ions in the ratio Li/Mo= 0.1 were added and equal amount of charges to the S atoms were re-introduced to maintain overall charge neutrality. We built two model systems comprising 1T' MoS$_2$ bilayers without (system A) and with (system B) Li-ions in between the MoS$_2$ layers and both the systems were solvated in water (Supplementary Fig. 28). In the simulation, Li/Mo ratio was kept at 0.1. A corresponding number of sulfur atoms in MoS$_2$ were charged with an additional electron to obtain a charge-neutral system. The initial d-spacings of the two MoS$_2$ bilayers were set to 6.4 Å. It is to be noted that system A is charge neutral.

Non-bonded interactions between atoms $i$ and $j$ were described by Lennard-Jones (LJ) and Coulombic potentials with the parameters $\varepsilon$, $\sigma$, and $q$ as follows:

$$U_{ij} = \sum_i \sum_j 4\varepsilon_{ij} \left[ \left( \frac{\sigma_{ij}}{r_{ij}} \right)^{12} - \left( \frac{\sigma_{ij}}{r_{ij}} \right)^{6} \right] + \frac{q_i q_j}{4\pi\varepsilon_0 r_{ij}}$$

where the Lorentz-Berthelot combining rules are applied to obtain $\varepsilon_{ij} = (\varepsilon_i \varepsilon_j)^{1/2}$ and $\sigma_{ij} = \frac{\sigma_i + \sigma_j}{2}$ $r_{ij}$ is the distance between atoms $i$ and $j$; $q_i$ and $q_j$ are the charges of atoms $i$ and $j$; $\varepsilon_0$ is the vacuum permittivity. The parameters of non-bonded interaction used in the simulations are presented in Supplementary table 2. The water molecules were described using the SPC/E model[29].

**Supplementary Table 2.** Non-bonded interaction parameters of 1T' MoS$_2$.

| $i$ | $\sigma_i$ (nm) | $\varepsilon_i$ (kJ mol$^{-1}$) | $q_i$ (e) |
|---|---|---|---|
| Li$^+$ | 0.2740 | -0.00233 | 1.0000 |
| Mo1 | 0.2862 | -0.13 | 1.0248 |
| Mo2 | 0.2862 | -0.13 | 1.0213 |
| S1 | 0.3928 | -0.25 | -0.4903 |
| S2 | 0.3928 | -0.25 | -0.5330 |

All simulations were performed with the NAMD package[30] (version 2.14). Periodic boundary conditions were applied in three dimensions to minimize the edge effect. Temperature and pressure were controlled by the Langevin dynamics and Langevin piston Nosé-Hoover methods[31], respectively. During the simulations, the bonds of each molecule were constrained to the nominal length using the SHAKE algorithm. Full electrostatic interaction was solved using the Particle Mesh Ewald (PME) model[32] with the grid spacing of 1.0 Å. Switching functions were used to smoothly truncate van der Waals, and electrostatic interactions to zero at the cut-off distance of 12 Å. A shifting function was adopted to separately solve the long and short electrostatic interactions with the cut-off distance. Each system was first equilibrated with the $MoS_2$ bilayer constrained at the initial position in the isothermal-isobaric (*NPT*) for 5.0 ns under atmospheric pressure, and 298 K. Subsequently, the constraint vertical to the $MoS_2$ plane was released to allow the bilayer to expand freely, and an additional simulation run for 5.0 ns was carried out in the canonical (*NVT*) ensemble to collect the swelling information of $MoS_2$ bilayer. The trajectory time step was 2 fs and data were sampled every picosecond.

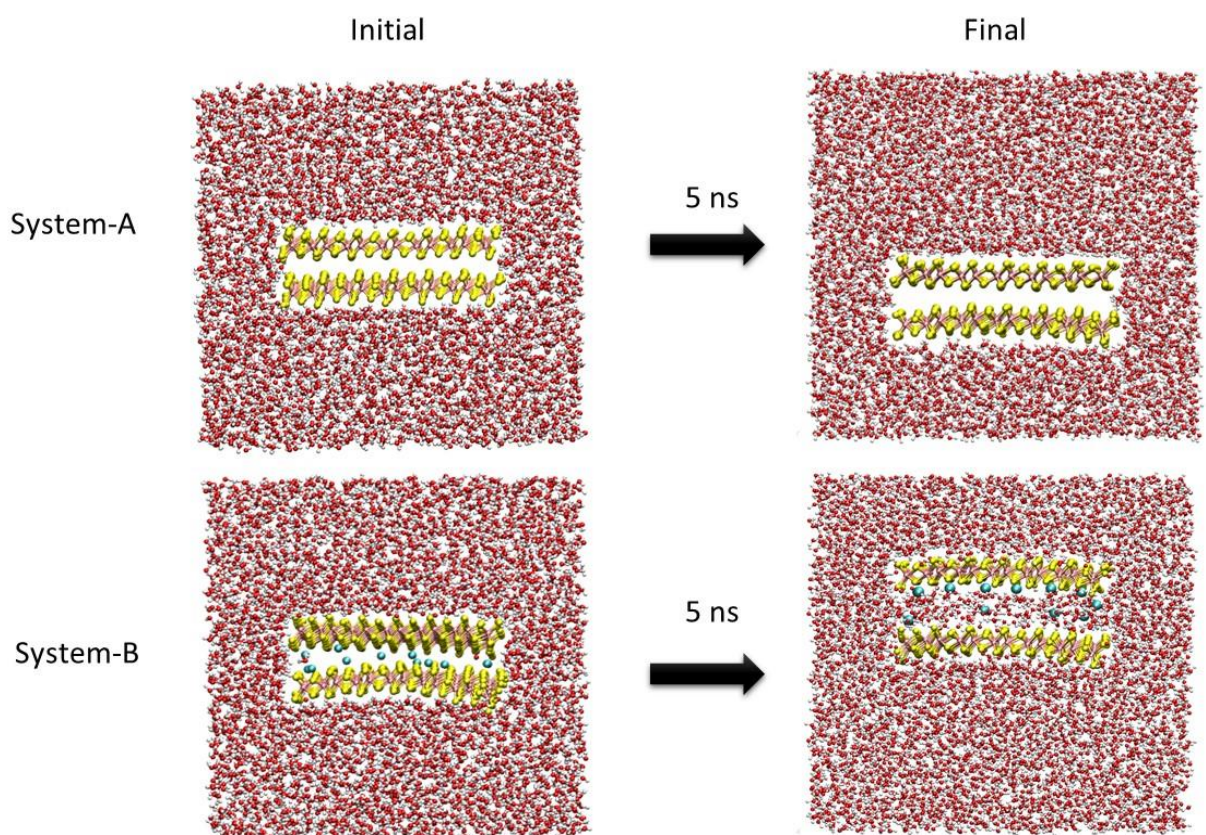

**Supplementary Fig. 28. Classical MD simulation – Effect of Li ions on water intercalation.** Snapshot after 5 ns simulation of unlithiated (system A) and $Li^+$ ion-containing (system B) $MoS_2$ bilayers in water.

Supplementary Fig. 28 shows snapshots of the $MoS_2$ bilayers (System A and B) in water before and after MD simulation of 5 ns. It is evident that system A maintains the initial structure with the *d*-spacing of 6.4 Å, while system B expands to 11.5 Å by incorporating water molecules in the interlayer space. This is in agreement with the experimental X-ray diffraction data shown in Supplementary Fig. 3a. Since the permeation of water through the $MoS_2$ laminate happens via the interlayer spaces, the absence of water permeation in 2H and the acid-treated $MoS_2$ laminate membrane can be explained by the absence of water intercalation into the interlayer channel.

To further understand the water intercalation process, we monitored the orientation of the water molecule dipole near the $MoS_2$ bilayer entrance. Supplementary Fig. 29 shows the water dipole orientation profile along the channel direction (X-axis) in the modelling system A. It is evident that water molecules nearest to the $MoS_2$ sheets have a stable and preferential orientation, with the hydrogen atoms facing the edge. This can be due to the uncompensated negative charges on sulphur edge atoms of the $MoS_2$ bilayer. Moving away from the $MoS_2$ sheets, the fluctuation of the water dipoles increases drastically, denoting the absence of any major long-range interaction acting on the water molecules. Within 15 Å from the edge of the sheets, the water molecules lose all of their preferential orientation, with the mean value of the water dipole moment orientation returning to 90 degrees.

Further, we looked into the dynamics of water molecules closest to the $MoS_2$ channel. Supplementary Fig. 30 shows the time dependence of the dipole orientation of the water molecules near the $MoS_2$ bilayer entrance of system A and system B. The smaller fluctuation of water dipoles in system A implies stable water configurations and hence the absence of intercalation. For system B, the water molecules near the bilayer entrance have a stable orientation at the initial stages of the simulation. However, with time, the configuration gradually fluctuates and becomes strongly disordered, indicating the entry of water molecules into the interlayer space.

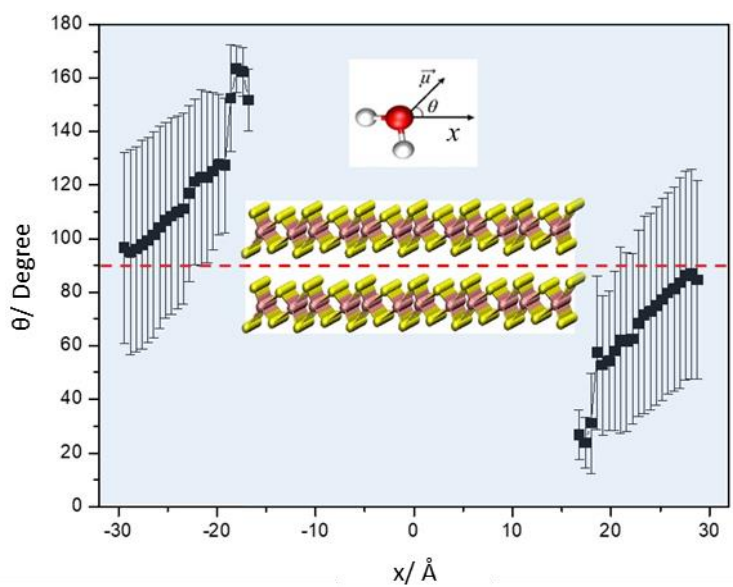

**Supplementary Fig. 29. Water dipole orientation.** Profile of water dipole orientation along the channel direction (X-axis) in modelling system A. The average dipole moment angle θ in a line along the channel direction is shown. The error bar along the Y-axis denotes the degree of fluctuation.

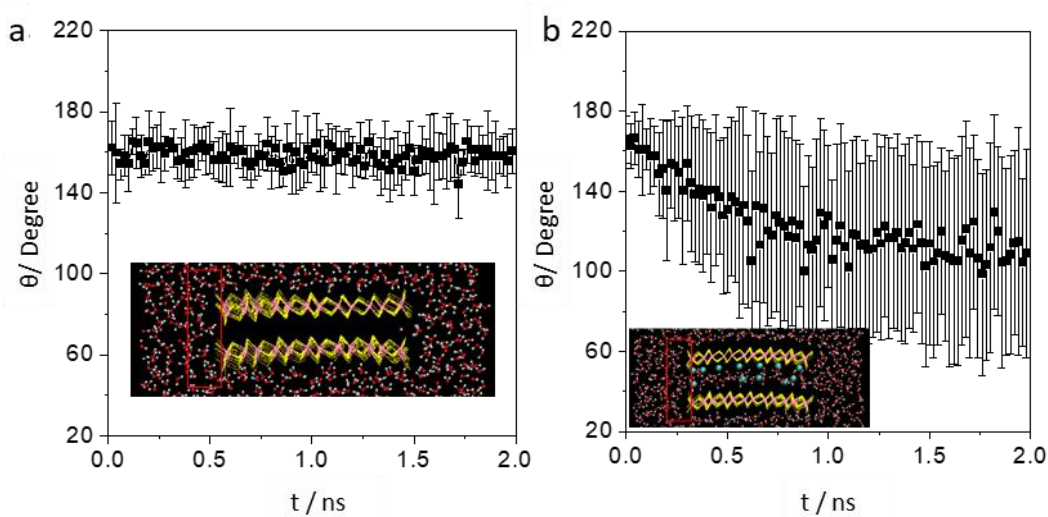

**Supplementary Fig. 30. Evaluation of water dipole moment near the MoS$_2$ bilayer entrance.** Dipole orientation of the water near the interlayer spacing (as marked in the inset) of system A (a) and system B (b) as a function of simulation time.